\documentclass[prd,aps,floatfix,preprint,nofootinbib]{revtex4}
\usepackage{graphicx}
\usepackage{amsmath}
\usepackage{amssymb}
\usepackage{longtable}

\def\drvstar#1{\partial\kern-0.5pt\smash{\raise 4.5pt\hbox{$\ast$}}
               \kern-5.0pt_{#1}} 
\def\lvec#1{\setbox0=\hbox{$#1$}
    \setbox1=\hbox{$\scriptstyle\leftarrow$}
    #1\kern-\wd0\smash{
    \raise\ht0\hbox{$\raise1pt\hbox{$\scriptstyle\leftarrow$}$}}
    \kern-\wd1\kern\wd0} 
 
\def\ldrvstar#1{\lvec{\,\partial}\kern-0.5pt\smash{\raise 4.5pt\hbox{$\ast$}}
               \kern-5.0pt_{#1}}
\newcommand{\simg}{\rlap{\raise -4pt \hbox{$\sim$}}
                   \raise 3pt \hbox{$>$}}
\newcommand{\siml}{\rlap{\raise -4pt \hbox{$\sim$}}
                   \raise 3pt \hbox{$<$}}
\newcommand{\del}{\partial}  
\newcommand{\no}{\nonumber}
\newcommand{\la}{\langle}
\newcommand{\ra}{\rangle}

\def\mpin{{m_{\pi^0}}}
\def\mpic{{m_{\pi^{+}}}}

\def\mKn{{m_{K^0}}}
\def\mKc{{m_{K^+}}}

\def\alpem{\alpha}

\def\mres{m_{\rm res}}

\def\mval{m_{\rm val}}
\newcommand{\mvf}{\mbox{$m_{\rm val,1}$}}
\newcommand{\mvs}{\mbox{$m_{\rm val,2}$}}
\def\msea{m_{\rm sea}}
\newcommand{\x}{{\mbox{$\vec{x}$}}}
\newcommand{\y}{{\mbox{$\vec{y}$}}}

\def\drvstar#1{\partial\kern-0.5pt\smash{\raise 4.5pt\hbox{$\ast$}}
               \kern-5.0pt_{#1}} 
\def\lvec#1{\setbox0=\hbox{$#1$}
    \setbox1=\hbox{$\scriptstyle\leftarrow$}
    #1\kern-\wd0\smash{
    \raise\ht0\hbox{$\raise1pt\hbox{$\scriptstyle\leftarrow$}$}}
    \kern-\wd1\kern\wd0} 
 
\def\ldrvstar#1{\lvec{\,\partial}\kern-0.5pt\smash{\raise 4.5pt\hbox{$\ast$}}
               \kern-5.0pt_{#1}} 
\begin{document}   
 \vspace*{-20mm}
 \begin{flushright}
  \normalsize
KANAZAWA-07-11\\
  KEK-CP-189 \\
  RBRC-636 
 \end{flushright}
\title{
 Determination of light quark masses from the electromagnetic splitting
 of pseudoscalar meson masses computed with two flavors of domain wall
 fermions
}
\author{
 Thomas~Blum$^{a,b}$,
 Takumi~Doi$^{b}$\footnote{Present address: Department of Physics and Astronomy, University of Kentucky, Lexington, KY 40506, USA.}
 Masashi~Hayakawa$^{c}$,
 Taku~Izubuchi$^{b,d}$,
 Norikazu~Yamada$^{e,f}$\\
 (for the RBC Collaboration)
}
\affiliation{
$^a$Physics Department, University of Connecticut, Storrs, CT
    06269-3046, USA\\
$^b$RIKEN-BNL Research Center, Brookhaven National Laboratory, Upton,
    NY 11973, USA\\
$^c$Department of Physics, Nagoya University, Nagoya 464-8602, Japan\\
$^d$Institute of Theoretical Physics, Kanazawa University, Kanazawa
    920-1192, Japan\\
$^e$High Energy Accelerator Research Organization {\rm (}KEK{\rm)}, Tsukuba
    305-0801, Japan\\
$^f$School of High Energy Accelerator Science, The Graduate University
    for Advanced Studies {\rm (}Sokendai{\rm )}, Tsukuba 305-0801, Japan
}
\date{\today}
\begin{abstract}
 We determine the light quark masses from lattice QCD simulations
 incorporating the electromagnetic interaction of valence quarks, using
 the splittings of charged and neutral pseudoscalar meson masses as
 inputs.
 The meson masses are calculated on lattice QCD configurations
 generated by the RBC Collaboration for two flavors of dynamical domain
 wall fermions, which are combined with QED configurations generated via quenched non-compact lattice QED.
 The electromagnetic part of the pion mass splitting is found to be
 $m_{\pi^+}-m_{\pi^0}=4.12(21)$ MeV, where only the statistical error is quoted, and similarly for the kaon, 1.443(55) MeV. 
 Our results for the light quark masses are
 $m_u^{\rm\overline{MS}}$(2 GeV)=$3.02(27)(19)$ MeV,
 $m_d^{\rm\overline{MS}}$(2 GeV)=$5.49(20)(34)$ MeV, and
 $m_s^{\rm\overline{MS}}$(2 GeV)=$119.5(56)(74)$ MeV, where the first error
 is statistical and the second reflects the uncertainty in our non-perturbative  renormalization procedure.  By averaging over $\pm e$ to cancel
 ${\cal O}(e)$ noise exactly on each combined gauge field configuration, we are able to work at physical $\alpha=1/137$ and
 obtain very small statistical errors.
 In our calculation, several sources of systematic error remain, including finite volume, non-zero lattice spacing, chiral extrapolation, quenched QED, and quenched strange quark, which may be more significant than the errors quoted above. We discuss these systematic errors and how to reduce or eliminate them.
\\[1ex]
PACS : 11.25.-w, 11.25.Yb, 11.30.Cp
\end{abstract}
\maketitle
%
\setcounter{footnote}{0}
\section{Introduction}
\label{sec:intro} 

Electromagnetic (EM) properties of hadrons offer a rich source of
interesting and important phenomena.
The patterns of the mass splittings between charged and neutral mesons
or the mass splittings among the octet or decuplet baryons are sensitive
to the isospin breaking from  different up and down quark masses and the
EM interactions.
It is also known that the width difference of $\rho^+$ and $\rho^0$ and
the off-shell hadronic light-by-light scattering amplitude, which can
not be measured experimentally, play an important role in the Standard
Model (SM) prediction of the anomalous magnetic moment of muon.

Recent developments made in both hardware and software have advanced
lattice QCD close to the goal of realistic QCD calculations, and large
scale, high precision unquenched simulations are becoming
available~\cite{Davies:2003ik,Izubuchi:lattice2005,giusti:lattice2006}.
Statistical errors on pseudoscalar meson masses are well under control,
with typical sizes of one-half of one percent, or less.
Recalling that charged-neutral meson mass splittings are of ${\cal
O}(\alpha)\sim O(1\%)$, where $\alpha$ is the fine structure constant,
it is expected that once EM interactions are successfully included, it
will be possible to determine the up and down quark masses from first
principles by using such splittings as inputs. Thus, among others, one
can check the simplest solution to the strong CP problem, $m_u$=0.

In this work, we focus on the determination of the light quark masses
using the charged-neutral splittings of the light pseudoscalar meson
masses. Following the pioneering work
in~\cite{Duncan:1996xy,Duncan:1996be}, we introduce EM fields in a
non-compact form and combine them with QCD gauge fields to realize the
combined QCD + QED theory on the lattice.
While these earlier works were done with unimproved Wilson quarks in the
quenched approximation of QCD, we employ domain wall fermions
(DWF)~\cite{Kaplan:1992bt,Shamir:1993zy} on configurations with two
flavors of dynamical quarks, generated by the RBC
Collaboration~\cite{Aoki:2004ht}, which allows us to realize the
calculation with less systematic errors.

Our calculation does not contain either effects of the dynamical strange
quark or the EM interactions of the dynamical quarks.
Removing the former approximation is now not an obstacle\footnote{
The RBC and UKQCD collaborations are generating 2+1 flavor domain wall
fermion ensembles of gauge fields, with even lower masses and larger
volumes than those used in the present study. We plan to use these in
future calculations of the kind reported on here.}.
A cost-effective method to avoid the latter approximation has been
proposed in~\cite{Duncan:2004ys}.
Since photon fields are massless and not confined, finite size effects
may be a significant source of error.
We examine these effects in the vector-saturation
model~\cite{Das:1967it,Bardeen:1988zw,Ecker:1988te,Donoghue:1993hj,Harada:2004qn}.
In addition to the above, we neglect disconnected quark-loop diagrams in
the pion two-point functions as the statistical errors associated with
them are difficult to control with current methods.
We will discuss this point in some detail in the following sections and
conclude that 
the omission of these diagrams does not affect the determination
 of light quark masses and has marginal effect on the determination
 of meson mass splittings.
 
We introduce the EM interactions on the lattice nonperturbatively such
that the effects include all orders of the $\alpem$ expansion.
Our analysis with several different values of $\alpem$ ranging from the
physical value $1/137$ to about $10/137$ shows that the $\alpem$
dependence of the splitting is well described by a linear fit over most of this 
range.
This fact encourages a perturbative introduction of the EM interactions
as an alternative~\cite{Hayakawa:2005eq}.

The rest of this paper is organized as follows.
We first give a brief overview on the combined QCD+QED lattice
calculation in Sec.~\ref{sec:qcd+qed system}.
The formula for the pseudoscalar meson mass in the presence of isospin
violation is given in Sec.~\ref{sec:mass formulas}.
After introducing the simulation details in
Sec.~\ref{sec:simulation details}, the numerical results and discussion
of systematic errors are given in Sec.~\ref{sec:results}, and
then concluding remarks are described in Sec.~\ref{sec:summary}.
Preliminary results of this work have been reported
in Ref.~\cite{Yamada:2005dv}, and our related activity toward the lattice
calculation of the anomalous magnetic moment of muon has been reported
in Ref.~\cite{Hayakawa:2005eq}.
Finally, we note that a similar calculation as the one given here, but
in the quenched approximation, has been reported
in Ref.~\cite{Namekawa:2005dr}.

\section{QCD+QED Calculations}
\label{sec:qcd+qed system}

In order to understand systematic errors in our lattice
study, we recall some basic material of the QCD+QED system.
In Sec.~\ref{subsec:globalSymmetryContinuum}, we discuss the
global symmetry of continuum QCD+QED before and after the spontaneous
breaking due to QCD dynamics.
In Sec.~\ref{subsec:disconnected}, how our approximation with which the
neutral pion is calculated affects its mass is discussed.
Sec.~\ref{subsec:WT_identity_in_DWF} deals with the axial Ward-Takahashi
identity in the QCD+QED system on the lattice with domain wall fermions.

\subsection{global symmetry in the continuum theory} 
\label{subsec:globalSymmetryContinuum} 

Here we identify the global symmetry of the continuum QCD+QED system and
the related Ward-Takahashi identity.
For now we set all quark masses to zero.

In pure QCD, the global symmetry of the fermion action is
$G_{\rm QCD} = SU(3)_L \times SU(3)_R \times U(1)_V$, where $U(1)_V$
represents the baryon number charge and $SU(3)_{L,R}$ are independent
flavor rotations on the left and right handed light quark fields,
respectively.
$U(1)_A$ does not exist due to the axial anomaly.
The symmetry which survives after spontaneous chiral symmetry
breaking due to QCD dynamics is the vectorial part of $G_{\rm QCD}$,
{\it i.e.}, $H_{\rm QCD} = SU(3)_V \times U(1)_V$. 

The QCD+QED system is obtained by introducing the EM interaction of
quarks $q = (u,\,d,\,s)^T$ with the $U(1)_{\rm em}$ gauge potential
$A_{{\rm em}\,\mu}$ as
\begin{eqnarray} 
 S_{\rm em} 
 &=& 
 \int d^4 x\, e j_{\rm em}^\mu\,A_{{\rm em}\,\mu} \, , \\ 
 j_{\rm em}^\mu
 &=& 
 \overline{q} \gamma^\mu Q_{\rm em} q \,,
\end{eqnarray} 
where $j_{\rm em}^\mu$ is the electromagnetic current, and
$Q_{\rm em}$ is the $3 \times 3$ matrix of the electric charges of the
quarks.
\begin{eqnarray} 
 && 
 Q_{\rm em} = 
 \left( 
  \begin{array}{ccc} 
   \frac{2}{3} & 0 & 0 \\ 
   0 & -\frac{1}{3} & 0 \\ 
   0 & 0 & -\frac{1}{3} 
  \end{array} 
 \right) \, . 
\end{eqnarray} 
It is convenient to express $Q_{\rm em}$ in terms of the generators
$\left\{T^a\right\}_{a=1,\,\cdots,\,8}$ of $SU(3)$, which can be given
by $T^a = \frac{\lambda^a}{2}$ with $\lambda^a$ the Gell-Mann matrices
and satisfy ${\rm tr}\left(T^a T^b\right) = \frac{1}{2} \delta^{ab}$, as 
\begin{eqnarray} 
 && 
 Q_{\rm em} = T^3 + \frac{1}{\sqrt{3}} T^8 \, .
\end{eqnarray} 
The EM interaction $S_{\rm em}$ breaks a part of $G_{\rm QCD}$.
This can be seen explicitly by looking at the modified Ward-Takahashi
identity for the flavor-nonsinglet axial-vector current,
$A^a_\mu \equiv \overline{q} \gamma_\mu \gamma_5 T^a q$,
\begin{eqnarray} 
 \del^\mu A^a_\mu
 = 
 i e A_{{\rm em}\,\mu}\, 
 \overline{q} \left[T^a,\,Q_{\rm em}\right] \gamma^\mu \gamma_5 q
 - 
 \frac{\alpha}{2\pi}\, 
 {\rm tr}\left(Q_{\rm em}^2 T^a\right)
 F_{\rm em}^{\mu\nu} \widetilde{F}_{{\rm em}\,\mu\nu} \, , 
  \label{eq:axialWT-massless}
\end{eqnarray} 
where
$F_{{\rm em}\,\mu\nu} = \del_\mu A_{{\rm em}\,\nu} - \del_\nu A_{{\rm em}\,\mu}$
is the field strength of the electromagnetic field,
$\widetilde{F}_{{\rm em}\,\mu\nu} \equiv \frac{1}{2} 
\varepsilon_{\mu\nu\lambda\rho} F_{\rm em}^{\lambda\rho}$,
and we still have the light quark masses set to zero.
The second term on the right hand side arises from the QED chiral
anomaly.
The first term on the right hand side vanishes for $a = 3,\,6,\,7,\,8$ 
in the Gell-Mann basis while the second term does so for
$a =1,\,2,\,4,\,5,\,6,\,7$, and a linear combination of $a =$ 3 and 8,
\begin{eqnarray} 
 \frac{1}{2}T^{\prime\,3} 
 &\equiv& 
 \frac{\sqrt{3}}{2} T^8 - \frac{1}{2}T^3 
 = 
 \frac{1}{2}
 \left( 
  \begin{array}{ccc} 
   0 & 0 & 0 \\ 
   0 & 1 & 0 \\ 
   0 & 0 & -1 
  \end{array} 
 \right) \, . 
\end{eqnarray} 
 Thus the global symmetry present in this system is 
\begin{eqnarray} 
 && 
 G_{\rm QCD + QED} 
 = SU(2)^\prime_L \times  SU(2)^\prime_R 
   \times U(1)_{\rm em} \times U(1)_V \, , 
\end{eqnarray} 
where $SU(2)^\prime_{L,\,R}$ are the subgroups of $SU(3)_{L,\,R}$
generated by $T^6,\,T^7,\,T^{\prime\,3}$.
QCD dynamics breaks $G_{\rm QCD + QED}$ spontaneously down to the
vector-like symmetry
\begin{eqnarray} 
 && 
 H_{\rm QCD + QED} 
 = 
 SU(2)^\prime_{V} \times U(1)_{\rm em} \times U(1)_V \, . 
\end{eqnarray}
The exact Nambu-Goldstone (NG) bosons associated with this spontaneous
symmetry breaking in massless QCD+QED are $K^0$, $\overline{K}^0$ and a
neutral meson corresponding to the generator $T^{\,\prime\,3}$,
{\it i.e.}, $d\bar{d} - s \bar{s}$.

Now remember that without the second term on the right hand side of
Eq.~(\ref{eq:axialWT-massless}) due to the QED anomaly, four neutral
mesons are massless in the chiral limit as
Eq.~(\ref{eq:axialWT-massless}) then vanishes for $a=$3, 6, 7, and 8.
In other words, the QED anomaly may make the meson corresponding to the
generator $Q_{\rm em}$ massive.
This meson acquires its mass through a diagram consisting of two quark
triangles connected by two photons, which is ${\cal O}(\alpha^2)$.
Since we neglect ${\cal O}(\alpem^2)$-effects throughout this work, we
will consider this meson as a NG boson up to this approximation.
Neglecting the QED anomaly term, the pattern of chiral symmetry breaking
in massless QCD+QED reads
\begin{eqnarray} 
 && 
 \widehat{G}_{\rm QCD + QED} 
 = 
 SU(2)^{\prime}_L \times U(1)^{\prime\prime}_L \times 
 SU(2)^{\prime}_R \times U(1)^{\prime\prime}_R \times U(1)_V  
  \nonumber \\
 && \quad 
 \Rightarrow 
 \widehat{H}_{\rm QCD + QED} 
 = 
 SU(2)^{\prime}_V \times U(1)_{\rm em} \times U(1)_V \, . 
  \label{eq:pattern_QEDanomalyNeglected}
\end{eqnarray} 
Here $U(1)^{\prime\prime}_{L/R}$ are the subgroups of $SU(3)_{L/R}$
generated by $Q_{\rm em}$, respectively.
The Goldstone bosons associated with this pattern are the strange mesons
$K^0$, $\bar{K}^0$ and $\pi_3$, $\eta_8$ corresponding to $T^3$, $T^8$.

\subsection{mass of the neutral pion} 
\label{subsec:disconnected}

In pure QCD, as long as $m_u=m_d$, isospin symmetry remains unbroken,
and we need consider only the connected (quark) diagram in the neutral
pion correlation function since the disconnected ones cancel
exactly\footnote{
The interpolating operator for $\pi_3$ is
$(\bar u\gamma_5 u -\bar d \gamma_5 d)/\sqrt{2}$, so in pure QCD the
four resulting disconnected diagrams in the correlation function
$\langle \pi_3\pi_3\rangle$ cancel. Once isospin symmetry is broken, the
up and down quark loops are no longer equal and hence do not cancel.}.
However, once the EM interaction and the $u$-$d$ mass difference are
introduced, disconnected diagrams no longer cancel and must be included.
Then two complications arise; one is the mixing of neutral pseudoscalar
mesons and the other is the appearance of the two-photon state as the
ground state\footnote{
 Since the two-photon state must carry orbital angular momentum, each
 photon has non-zero momentum.
 With our lattice setup, the possible minimum energy for this state is
 about 1.3 GeV.
 Therefore we could have neglected this state even if we had included
 disconnected diagrams since the possible minimum energy is much larger
 than the NG boson masses.
}.
Due to the well known computational difficulty associated with
disconnected diagrams, we neglect their contributions everywhere in our
current study.
Thus we need to know how this approximation affects the pion mass.
Remember that calculating the connected diagram only is equivalent
to calculating the mass of the neutral kaon, and the neutral kaon is
massless in the chiral limit in the presence of the EM interaction as
discussed in Sec.~\ref{subsec:globalSymmetryContinuum}.
Therefore the neutral pion mass computed with only connected diagrams
does not have a term of $O(\alpem)$ in the chiral limit.

For clarity, let $\pi_3$ and $\pi^0$ denote the neutral pion in the
basis of flavor eigenstates and mass eigenstates, respectively.
In Sec.~\ref{subsec:globalSymmetryContinuum}, we discussed that including terms of
$O(\alpem)$, the neutral pion corresponding to $T^3$, {\it i.e.}
${\pi_3}$, is massless when $m_u=m_d= 0$.
In general, the mass of the physical neutral pion, $m_{\pi^0}$, is
different from $m_{\pi_3}$ due to the mixing with other neutral
pseudoscalar mesons.
Following the classic current algebra treatment of $\pi_3-\eta$
mixing~\cite{Gross:1979ur} and the $\pi_3-\eta^\prime$
mixing~\cite{Gasser:1984gg} based on ChPT or a version of that,
in pure QCD with $m_u\ne m_d$ the contribution to $m_{\pi^0}^2$ is
estimated to be proportional to $(m_u-m_d)^2 /m_s^2$.
Repeating a similar treatment for the EM interaction,
it turns out that the fact that $m_{\pi^0}^2$, including $O(\alpem)$ corrections, vanishes in the chiral
limit does not change.
Hence, the mass-squared of $\pi^0$ begins at second order in the isospin
breaking $(m_u-m_d)^2$ and/or $\alpem^2$.
Since we neglect effects of this order throughout this work, we can
identify ``the neutral pion'' consisting of only connected diagrams with
that of $\pi^0$.

In summary, as long as we discuss the squared pseudoscalar meson mass
and neglect ${\cal O}(\alpem^2)$ and ${\cal O}((m_u-m_d)^2)$-effects,
``the neutral pion mass'' in this work is approximately equal to the
physical $\pi^0$ mass, and the difference can arise at
${\cal O}(\alpem m)$.

\subsection{axial Ward-Takahashi identity with domain wall fermions}
\label{subsec:WT_identity_in_DWF} 

We consider the QCD+QED system on the lattice with two flavors of domain
wall quarks, $q_1$ and $q_2$, which have mass $m_{\rm val,1}$ and
$m_{\rm val,2}$ and charge $Q_{\rm val,1}$ and $Q_{\rm val,2}$,
respectively.
These fields are valence quark fields. 
The flavor non-singlet axial Ward-Takahashi (WT) identity associated
with the chiral transformation of the valence quark fields is given by
\begin{eqnarray}
 \partial_{\mu}^* {\cal A}^a_\mu(x)
 &=&
 (m_{\rm val,1} + m_{\rm val,2})\,P^a(x)
 + (m_{\rm val,1} - m_{\rm val,2})\,\bar{q}(x) \gamma_{5} 
   \left\{ 
    \frac{\tau^a}{2},\frac{\tau^3}{2} 
   \right\}\,q(x)
  \nonumber \\ 
 && \ 
 + 2\,J_{5q}^a(x)
 + \sum_s \epsilon(s){\cal X}_s^a(x) \, ,
 \label{eq:latt pcac}
\end{eqnarray}
where $q=\left(q_1,q_2\right)^T$, 
${\cal A}_\mu^a(x)$ is the conserved axial-vector current whose
form is the same as that in the pure QCD domain wall fermion
system~\cite{Furman:1994ky}, but the link variables are now replaced
with
\begin{eqnarray} 
 \left( 
  \begin{array}{cc} 
   U_{{\rm QCD},\,\mu}(x) 
    \left(U_{{\rm em},\,\mu}(x)\right)^{e Q_{\rm val,1}} & 0 \\  
   0 & U_{\rm QCD,\,\it\mu}(x) \left(U_{{\rm em},\,\mu}(x)\right)^{e
    Q_{\rm val,2}} 
  \end{array}  
 \right) \,.
\end{eqnarray}
$P^a(x)$ and $J_{5q}^a(x)$ denote pseudoscalar densities.
While $P^a(x)$ contains physical quark fields only, $J_{5q}^a(x)$
is defined at the midpoint of the fifth dimension and written in terms
of bulk fields, as in the pure QCD case~\cite{Furman:1994ky}.
${\cal X}_s^a(x)$ is give by
\begin{eqnarray}
 {\cal X}_s^a(x)
 &=& 
 -\frac{1}{2} 
 \sum_\mu 
 \left[ 
  \bar\Psi_s(x) (1-\gamma_\mu) 
  U_{{\rm QCD},\,\mu}(x) 
 \right.
  \nonumber \\
 && \qquad \qquad
  \times
  \left( 
     (U_{\rm em,\it\mu}(x))^{eQ_{\rm val,1}} 
   - (U_{\rm em,\it\mu}(x))^{eQ_{\rm val,2}}
  \right) 
   \left[\frac{\tau^a}{2},\frac{\tau^3}{2}\right]
    \Psi_{s}(x + \widehat{\mu})
  \nonumber \\
 && \qquad \quad
   + \bar\Psi_s(x)
     (1 + \gamma_\mu) U^\dag_{{\rm QCD}\,\,\mu}(x - \widehat{\mu})\,
   \nonumber \\
 && \qquad \qquad
  \times
      \left( 
       \left(U^{\dagger}_{{\rm em}\,\,\mu}(x -
        \widehat{\mu})\right)^{eQ_{\rm val,1}}
        - \left(U^{\dagger}_{{\rm em},\,\mu}(x -
           \widehat\mu)\right)^{eQ_{\rm val,2}}
      \right) 
   \nonumber \\ 
  && \qquad \qquad
  \left. 
   \times 
    \left[\frac{\tau^a}{2},\frac{\tau^3}{2}\right] \Psi_s(x - \widehat{\mu}) 
  \right],
\label{eq:X}
\end{eqnarray}
where the lattice spacing is set to $1$, 
$\epsilon(s)=1$ for $0\le s \le L_s/2-1$ and $-1$ for
$L_s/2 \le s < L_s$. 
${\cal X}_s^a(x)$ vanishes when $a=3$, $e=0$ or
$Q_{\rm val,1}=Q_{\rm val,2}$, so it is the lattice counterpart to the
first term in Eq.~(\ref{eq:axialWT-massless}).
The formula analogous to Eq.~(\ref{eq:latt pcac}) is written down for
three flavor case.
From the analogy to the pure QCD case in the domain-wall
formalism~\cite{Shamir:1993yf}, it is inferred that the EM chiral
anomaly arises from $J^a_{5q}$ only after sandwiching
Eq.~(\ref{eq:latt pcac}) between physical states.
Due to the presence of the EM chiral anomaly, ${\cal A}_\mu^a(x)$ is not
conserved for $e Q_{\rm val,1}\ne e Q_{\rm val,2}$ even in the chiral
limit, and hence no NG boson shows up after spontaneous chiral symmetry
breaking in this two-flavor theory, only pseudo-NG bosons do.
However, since we neglect the ${\cal O}(\alpem^2)$ contributions in this work, the neutral pion can be regarded as a NG boson.

\section{formulas for pseudoscalar meson masses and splittings}
\label{sec:mass formulas}

\subsection{QCD with $m_u=m_d$}

First we remind the reader of the next-to-leading order (NLO) partially
quenched chiral perturbation theory (PQChPT) formula for the
``kaon-like'' pseudoscalar meson mass-squared in pure QCD.
In this formula, all $N_f$ sea quarks have degenerate mass $\msea$ and
valence quarks have masses set to
$\mvf = m_u = m_d = m_{ud}
\ne \mvs =m_s$~\cite{Golterman:1997st,Laiho:2002jq}.
\begin{eqnarray}
      m_{K\rm}^2
&=&   M_{K}^2 \left(1 + \frac{\Delta_{\rm NLO}(m_{K}^2)}{M_{K}^2}
              \right),
\label{eq:chpt 1-loop}\\
     \frac{\Delta_{\rm NLO}(m_{K}^2)}{M^2_{K}}
&=&   \frac{-1}{N_f(M^2_{K}-M^2_{\pi})}
      \left[   (M^2_{\pi}-M^2_{ss})A_0(M^2_{\pi})
             + (-M^2_{33}+M^2_{ss})A_0(M^2_{33}) \right]\no\\
& & - \frac{16}{f^2}\bigg[   (L_5-2L_8)M^2_{K}
                           + (L_4-2L_6)N_f\,M^2_{ss}\bigg],
\label{eq:chpt non-degenerate 1-loop}\\
      M^2_{\pi} &=& 2\,B_0\,m_{ud},
\label{eq:chpt mpi}\\
      M^2_{K}   &=& B_0\,(m_s + m_{ud}),
\label{eq:chpt mK}\\
      M^2_{33} &=& 2\,M^2_K - M^2_\pi,\\
      M^2_{ss} &=& 2\,B_0\,\msea,
\label{eq:chpt mss}\\
     A_0(M^2)
 &=& \frac{M^2}{16\pi^2f^2}
     \ln\left(\frac{M^2}{\Lambda^2_\chi}\right).
\end{eqnarray}
In the above $\Lambda_\chi$ denotes the renormalization scale of the
effective theory, $f$ the decay constant in the chiral limit, and $L_i$
the Gasser-Leutwyler low energy constants at $\Lambda_\chi$ that appear
in the ${\cal O}(p^4)$-chiral Lagrangian of QCD.
In the above, $m_P$ denotes the physical (one-loop) mass while $M_P$
denotes the tree level mass.
We take $N_f = 2$ according to our ensemble of gauge configurations.

In the limit $m_s\rightarrow m_{ud}$, the formula for the NLO
contribution to the mass-squared of ``pion-like'' mesons made of
degenerate valence quarks is obtained from the
above~\cite{Golterman:1997st,Laiho:2002jq},
\begin{eqnarray}
    \frac{\Delta_{\rm NLO}(m^2_{\pi})}{M^2_{\pi}}
&=& \frac{2}{N_f}\left[   \frac{ M^2_{\pi}-M^2_{ss}}{16\pi^2f^2}
                        + \frac{2M^2_{\pi}-M^2_{ss}}{M^2_{\pi}}
                          A_0(M^2_{\pi}) \right]\no\\
& & - \frac{16}{f^2}\bigg[(L_5-2L_8)M^2_{\pi}
    + (L_4-2L_6)N_f\,M^2_{ss}\bigg].
\label{eq:chpt degenerate 1-loop 2}
\end{eqnarray}

\subsection{$m_u \ne m_d$ effects}

Incorporation of the isospin breaking effects due to $m_u \ne m_d$ into the above mass formula is straightforward.
For kaons, this is achieved by simply replacing $m_{ud}$ in
Eq.~(\ref{eq:chpt mK}) with $m_u$ for $K^\pm$ or $m_d$ for $K^0$.
As for pions, $m_{ud}$ is replaced in Eq.~(\ref{eq:chpt mpi}) with
$(m_u+m_d)/2$.
At this order in ChPT a term proportional to $(m_u-m_d)^2$ also
appears~\cite{Gasser:1983yg}, {
but arises from the disconnected diagrams that contribute to the $\pi^0$ and $\eta$ correlation functions in QCD which we do not compute. So, this term is omitted from
our fits. 
}
\subsection{QED corrections to meson masses}
\label{subsec:qed correction}

{ At leading order, the correction to the charged meson mass 
squared is \cite{Dashen:1969eg},
\begin{eqnarray}
&\delta (Q_{1}-Q_{2})^2,&
 \label{eq:alp lo}
\end{eqnarray}
where $Q_{i}$ is the charge of
 {\it valence} quark $i$ in units of $e$.}
Since the meson masses do not change under simultaneous interchange
of quark mass and charge, the only possibilities for
${\cal O}(\alpem m)$-terms are
\begin{eqnarray}
&&(Q_{1}+Q_{2})^2(m_{1}+m_{2}),
 \label{eq:alp m 1}\\
&&(Q_{1}-Q_{2})^2(m_{1}+m_{2}),
 \label{eq:alp m 2}\\
&&(Q_{1}^2 - Q_{2}^2)(m_{1}-m_{2}),
 \label{eq:mu-md term}
\end{eqnarray}
{ where $m_{i}$ is the mass of
{\it valence} quark $i$}.
Combining this and the discussion in Sec.~\ref{sec:qcd+qed system}, we parameterize the EM contribution to the meson masses as
\begin{eqnarray}
\label{eq:emcorrection}
    \Delta_{\rm em}(m_{ij}^2)
  &=&
     \delta\,(Q_i-Q_j)^2\\\nonumber &+&
      \delta_{0}\,(Q_i+Q_j)^2\,(m_i+m_j)\\\nonumber &+&
      \delta_{+}\,(Q_i-Q_j)^2\,(m_i+m_j)\\\nonumber &+&
      \delta_{-}\,(Q_i^2-Q_j^2)\,(m_i-m_j)\\\nonumber &+&
      \delta_{\rm sea}\,(Q_i-Q_j)^2\,(2 \,m_{sea})\\\nonumber &+&
      \delta_{\rm m_{res}}\,(Q_i+Q_j)^2.
\end{eqnarray}
{ To avoid confusion, $Q_i$ always refers to the charge of the quark, not the anti-quark. Factors of $\alpha$ and $B_0$  have been absorbed into the definition of
the low energy constants ($\delta$'s)}.
${\cal O}(\alpem^2)$ contributions have been neglected in Eq.~(\ref{eq:emcorrection}). { The last LEC is purely a lattice artifact induced by the finite size of the extra dimension for DWF and vanishes in the limit $L_s\to \infty$. There is also a similar small mixing with the physical term, $\delta(Q_i-Q_j)^2$, which can be subtracted by measuring the EM effects on the residual mass. We will come to this point again when we discuss the residual mass and results of our fits in Section~\ref{sec:results}.}
The logarithmic dependence on the quark mass has been calculated for the
unquenched theory~\cite{Urech:1994hd} and very recently for the
partially-quenched theory~\cite{Bijnens:2006mk}, to NLO, {\it i.e.,}
including all terms of ${\cal O}(\alpha m)$ for the case of three
valence ($n_{val}=3$) and three sea ($n_{sea}=3$) quarks.
However such an analysis is not available yet for $n_{sea} = 2$;
we thus omit such terms. 
A full treatment of the splittings to NLO must await our upcoming
calculation with 2+1 flavors of sea quarks. 

{ Note that the masses of the charged mesons $\pi^+$ ($u \bar d$) and $K^+$ ($u\bar s$) do not vanish for massless quarks, and the only terms that contribute to the neutral mesons are the ones with $\delta_0$ and
$\delta_{m_{res}}$.

$\delta_{0}$, $\delta_{+}$, $\delta_{-}$, and $\delta_{\rm sea}$ represent linear
combinations of low energy constants (LEC's) in the
${\cal O}(\alpha p^2)$-part of the chiral Lagrangian ({\it e.g.}, see
Ref.~\cite{Bijnens:2006mk}).
They are determined by fitting the numerical data to the form
given in Eq.~(\ref{eq:emcorrection}).
However, since our neutral pion does not contain disconnected diagrams,
the value of $\delta_{0}$ extracted in this work,
which we take to be the same for all mesons,
could be different from the physical one for the $\pi^0$ and $\eta$ mesons.}

With the above parameterization, it is easy to see that Dashen's
theorem~\cite{Dashen:1969eg}
approximately holds, and that it is violated at
${\cal O}(\alpem\,m)$ which we parameterize as
\begin{eqnarray}
\Delta_{\rm EM} &=& \left(\frac{\mKc^2 -\mKn^2}{\mpic^2-\mpin^2}\right)_{\rm EM~part} - 1,
\label{eq:dashen break}
 \end{eqnarray}
following Ref.~\cite{Aubin:2004fs}.
Had we kept terms of ${\cal O}(\alpem^2)$, they would also violate
Dashen's theorem.

{ Adding the above contributions from QCD and QED, the squared
pseudoscalar meson masses are obtained at NLO,
\begin{eqnarray}
 m_{ij}^2 = M_{ij}^2 + \Delta_{\rm NLO}(M_{ij}^2) + \Delta_{\rm em}(M_{ij}^2).
 \label{eq:nlo pion mass}
\end{eqnarray}
}

\section{Simulation details}
\label{sec:simulation details}

We employ QCD gauge configurations with two flavors of dynamical domain
wall fermions, generated by the RBC Collaboration~\cite{Aoki:2004ht}.
The lattice size is $L^3\times T=16^3\times32$ with degenerate sea quark
masses, $a \msea=0.02$, $0.03$, $0.04$ in lattice units, corresponding to $m_s/2 \le
m_{sea} \le m_s$ where $m_s$ is the physical value of the strange quark mass.
In~\cite{Aoki:2004ht} the lattice spacing is determined to be 1.691(53)
GeV using $m_\rho$=770 MeV, and hence the physical spatial volume
corresponds to $V\approx(1.9 {\rm\, fm})^3$.
{
Most of the results presented below were determined on the same ensembles used in \cite{Aoki:2004ht}, about 100 lattices at each sea quark mass, separated by 50 monte-carlo time units (see Tab.~\ref{tab:ensembles}).
The source time of quark propagators is set to $t_{\rm src}=0$.
Additional information on the configuration ensemble is given in
Tab.~\ref{tab:qcd parameters}}.
For further details, see~\cite{Aoki:2004ht}.

A non-compact form is adopted for the lattice QED gauge action as
in~\cite{Duncan:1996xy}.
We first write the action in momentum space, and impose the Coulomb
gauge fixing condition { plus an additional constraint on the vector potential in the time direction (see Appendix A)}.
After diagonalizing the kernel, the Boltzmann weight for the photon
fields can be written in a Gaussian form.
We then generate them by the random Gaussian noise method with $e$=1.
The photon field in configuration space, $A_{\rm em,\it\mu}(x)$, is then
obtained by inverse Fourier transformation.
Details of the generation of the QED gauge configurations are given in
Appendix~\ref{app:gauge fixing}.
It is worth noting that, thanks to the non-compact form of the action
and thus the simple generation procedure, there is no auto-correlation
among the configurations.
Since we make the quenched approximation for QED, the sea quarks do not
couple to photons, and so the fine structure constant $\alpem$ does not
run with the cutoff scale.

Exponentiating $A_{\rm em,\it\mu}(x)$ and the valence quark's electric
charge $e\,Q_{\rm val}$, we obtain $U(1)$ link variables,
\begin{eqnarray}
    (U_{\rm em,\mu}(x))^{e\,Q_{\rm val}}
  = e^{i e\,Q_{\rm val} A_{\rm em,\mu}(x)}.
\end{eqnarray}
Configurations for the QCD+QED theory are then constructed from
$U_{\rm qcd,\it\mu}(x)\times (U_{\rm em,\it\mu}(x))^{e\,Q_{\rm val}}$,
which are used in the inversion of the Dirac operator for valence
quarks.

We calculate the two-point correlation functions of the pseudoscalar and
vector mesons with { seven valence quark masses, $\mval=0.015$, 0.020,
0.025, 0.030, 0.035, 0.040, and 0.0446 at each sea quark mass and compute all possible degenerate and non-degenerate two-point meson correlation functions. The last mass corresponds to the bare strange quark mass as determined in~\cite{Aoki:2004ht}}.
The electric charges of the valence quarks are $Q_{u}=+2/3$ for
up-type quarks and $Q_{d}=Q_{s}=-1/3$ for down-type quarks.
{ In a preliminary study
we took three values of $e$, which correspond to
$\displaystyle \alpha = 1/137,\ (0.6)^2/(4\pi),\ 1^2/(4\pi)$, to examine
the $\alpem$ dependence of the meson mass splittings. Having found that the relative statistical errors on our splittings did not change with $\alpha$, we fixed $\alpha$ to its physical value for the main calculations reported here. In addition, we found that statistical errors
in the splittings are dramatically reduced by averaging correlation functions with $e=\pm1$ as then the leading ${\cal O}(e)$ noise, which vanishes in ensemble averages, cancels exactly on each configuration. The dramatic reduction in error is shown in Fig.~\ref{fig:e+-1 trick} for a representative case.}

{ The lattice spacing determined from the charged vector meson  mass is found to be $a^{-1}$ = 1.659(37) GeV, using the experimental value of $m_\rho$= 775.8 MeV. This is consistent with
the value 1.691(53) GeV found in~\cite{Aoki:2004ht}. Our statistical error is smaller because in this case we used an ensemble of configurations twice the size used in~\cite{Aoki:2004ht}, and for most of the pseudoscalar mass splittings reported on here, by taking lattices separated by 20 and 25 monte-carlo units. These were
blocked together in pairs to account for possible auto-correlations.
}

Finally, we remind the reader that we have only calculated connected
diagrams for the neutral mesons.

\section{Numerical results}
\label{sec:results}
\subsection{residual quark mass}
\label{subsec:residual quark mass}

{
We first evaluate the residual quark mass in the usual way, using the midpoint
pseudoscalar density in Eq.~(\ref{eq:latt pcac}).
Since the residual quark mass depends on the EM charge, we determine it
separately for $\bar u u$, $d \bar d$, and $u \bar d$ mesons  by averaging the following ratio
over a suitable plateau for each type of meson~\cite{Blum:1998ud,Blum:2000kn},
\begin{eqnarray}
   a\mres^{a}(am_q)
 &= & \frac{\la J^a_{5q}(t) P^a(0) \ra}{\la P^a(t) P^a(0) \ra},
 \label{eq:mres_dwf}
\end{eqnarray}
where $O(t)$ denotes the operator $O(x)$ summed over the three-volume. 

In terms of a low energy chiral expansion, the ratio of correlation functions in Eq.~(\ref{eq:mres_dwf}) is a constant at LO and receives corrections at higher order
(proportional to $m_q$, $m_q^2$, and so on). We therefore define the
residual quark mass $m_{res}$ to be the value of the ratio in  Eq.~(\ref{eq:mres_dwf}) when $m_q\to 0$. 
The chiral limit is then defined as $m_{q}\to -m_{res}$. A similar strategy works here, but the definition of explicit chiral symmetry breaking effects in the low energy effective theory must be extended to include ${\cal O}(\alpha)$ effects.
Because the electric charges break isospin, the most general form
for the residual mass is
\begin{eqnarray}
m_{res}(\alpha) &=& m_{res}(m_q) + C_1 (Q_1-Q_2)^2 + C_2 (Q_1+Q_2)^2
\label{eq:mres em}
\end{eqnarray}
where $m_{res}(m_q)$ is the ratio in Eq.~(\ref{eq:mres_dwf}) with $\alpha=0$ and contains corrections to all orders in $m_q$, and
$C_1$ and $C_2$ are coefficients of ${\cal O}(\alpha)$ which vanish when $L_s\to \infty$. 

In fact, what matters in the following is the difference $\Delta m_{res} = m_{res}(\alpha) -m_{res}(m_q)$ which enters in mass-squared differences of the mesons. Because this difference is calculated from highly correlated values of the residual mass, it can be determined very precisely in each case. For $u \bar u$, $d \bar d$, and $u \bar d$ type mesons, we find
\begin{eqnarray}
\Delta m_{res,u \bar u} &=& 7.11(5)\times 10^{-5}\\
\Delta m_{res,d \bar d} &=& 1.73 (1)\times 10^{-5}\\
\Delta m_{res,u \bar d} &=& 2.36(4)\times 10^{-5},
\end{eqnarray}
where we have simply averaged the differences for all quark mass combinations. Notice that the ratio of the $u \bar u$ to $d \bar d$ value is very close to 4 as it must be according to Eq.~(\ref{eq:mres em}). However, 
$\Delta m_{res,u \bar d}$ is not equal to the average of the $u \bar u$ and $d \bar d$
values. Using $\Delta m_{res,u \bar d}$ and either the $u \bar u$ or $d \bar d$ value,
we can determine $C_1$ and $C_2$. We find
\begin{eqnarray}
C_1 &=& 1.92 \times 10^{-5}\\
C_2 &=& 4.00 \times 10^{-5}.
\end{eqnarray}
We ignore the tiny statistical errors in $C_1$ and $C_2$ in what follows below, and note
that the $\alpha=0$ value of $m_{res}$ determined here, 0.001387(39), is consistent with that found in \cite{Aoki:2004ht}.
}

\subsection{correlation function}
\label{subsec:ps meson mass}

To extract the pseudoscalar meson masses, we measure the following
two-point correlation function and fit it according to the same
procedure detailed in~\cite{Aoki:2004ht},
\begin{eqnarray}
     C_2^{P^{ij}}(t)
 &=& \la\,\left(P^{ij}(t)\right)^\dag\,P^{\rm wall,\it ij}(0)\,\ra
     \label{eq:2ptP},
\end{eqnarray}
where
\begin{eqnarray}
     P^{ij}(t)
  = \sum_\x \bar q_i(t,\x)\,\gamma_5\,q_j(t,\x),&\ \ &
     P^{\rm wall,\it ij}(t) 
  = \sum_{\x,\y} \bar q_i(t,\x)\,\gamma_5\,q_j(t,\y).
\end{eqnarray}

{ We consider the $n_{val}=3$ case, so
$q(t,\x)=(\,u(t,\x),\,d(t,\x),\,s(t,\x)\,)^T$, and we calculate the
above correlation functions for all possible combinations of degenerate and non-degenerate valence quark masses.
The pion in pure QCD is also calculated, and denoted by $\pi^{\rm Q}$.
The pseudoscalar meson masses computed from the pseudoscalar two-point functions are summarized in
Tabs.~\ref{tab:ps masses}~-~\ref{tab:ds masses}.
The fit range is the same in each case and for all quark masses, 9-16, and matches that used in~\cite{Aoki:2004ht}.
The value of $\chi^2$/dof for each fit is less than or about 1.
}

We calculate the vector meson masses in a similar manner.
{ The results for degenerate u,d quark mass mesons are summarized in Tab.~\ref{tab:vec masses}.
The quark mass dependence of the charged vector meson is shown in
Fig.~\ref{fig:vec_mass} as an example, where the line denotes the fit to
\begin{eqnarray}
   m_{\rho^\pm}
 = a_v + 2\,b_v\, \big( m_q+ m_{res,ud} \big),
 \label{eq:vec1}
\end{eqnarray}
using only the $m_q=\msea =\mval$ data points.
Figure~\ref{fig:vec_massdiff} shows the quark mass dependence of
the $\rho^\pm$-$\rho^0$ splitting where the $\msea =\mval$ points are fit to
\begin{eqnarray}
   m_{\rho^\pm}-m_{\rho^0}
 = c_v + 2\,d_v\, m_q,
 \label{eq:vec2}
\end{eqnarray}
and yield a small but nonzero value in the chiral limit ($\sim 0.5$ MeV).
However, it is seen from Fig.~\ref{fig:vec_massdiff} that if the chiral extrapolation is
made in a different way, for example by extrapolating $m_{\rm val}$ to
the chiral limit first, a different result is obtained.  It appears some of the
instability originates with the choice of fit range for the vector mass.
In Fig.~\ref{fig:vec_massdiff tmin5} we show the splitting calculated for a uniform $t_{min}=5$ for all sea quark masses. The spread in the splittings has cleary decreased. We note that the lower values of $t_{min}$ do not yield good $\chi^2$ for some of the $m_{sea}=0.03$ and 0.04 data, however, so there may be some excited state contamination in the splittings. In any case, it is clear that more statistics is needed to resolve the vector splitting, and in particular
its quark mass dependence.} 
Furthermore we have omitted disconnected diagrams in the calculation of
$m_{\rho^0}$, thus the nonzero value for splitting is not conclusive.
Note however, that at large $\mval$ where the statistical errors are
under better control, for all values of $\msea$ the EM splitting
corresponds to about 0.5 MeV.
Experimentally, the $\rho^+-\rho^0$ mass difference is consistent with
zero~\cite{Yao:2006px}.
The results for the parameters appearing in Eqs.~(\ref{eq:vec1}) and
(\ref{eq:vec2}) are tabulated in Tab.~\ref{tab:vector}.
The vector mass splitting is an interesting case for further study.

\subsection{size of ${\cal O}(\alpem^2)$-corrections}

Before going to the determination of the quark masses, let us discuss
the size of ${\cal O}(\alpem^2)$-corrections.
In Fig.~\ref{fig:e-dep of splittings} we show the $\alpem$ dependence of
$m_{\pi^+}^2-m_{\pi^0}^2$ and $m_{\pi^0}^2-m_{\pi^Q}^2$.
{ Three values of
the electric charge corresponding to $\alpem$=1/137, $0.6^2/(4\pi)$ and $1^2/(4\pi)$ are examined}.
The lines denote a linear fit to each set of mass-squared splittings
with the constraint that the splitting vanishes at $\alpem=0$.

While $m_{\pi^0}^2-m_{\pi^Q}^2$ is well described by the linear fit over
the whole range of $\alpha$, $m_{\pi^+}^2-m_{\pi^0}^2$ at the largest
value is clearly away from the fit line.
In both cases the data points for $\alpem$=1/137 and $0.6^2/(4\pi)$ lie
on a line including $\alpha=0$ within tiny statistical errors.
This indicates ${\cal O}(\alpem^2)$ effects only become significant for
$e\sim 1$, which is not terribly surprising.

\subsection{low energy constants and quark masses}

Now we turn to the determination of the quark masses.
To this end, we extract the low-energy constants (LEC's) defined in Sec.~\ref{subsec:qed correction} by fitting the
{ difference of the
square of the pseudoscalar mass with the same quantity evaluated at $\alpha=0$. 
The fits are uncorrelated since including the full covariance matrix
makes them unstable, and it is likely ill-determined besides.
Fitting all of the data at our disposal (61 masses at each sea quark mass), we find the LEC's tabulated in Tab.~\ref{tab:lec}. The mass squared differences and fit for $m_{sea}=0.02$ are shown in Fig.~\ref{fig:ps splittings}. Restricting the fit range to $m_f \le 0.03$ does not change the values significantly as seen in Tab.~\ref{tab:lec},
though the $\chi^2$/dof value is reduced. 
The corresponding $\alpha=0$ values of $B_0$ and the
NLO Gasser-Leutwyler constants are given in Tab.~\ref{tab:alpha=0 lec}.

The LEC's in Eq.~(\ref{eq:emcorrection}) are extracted using the
mass dependence of the form $m_i + m_{res}$. Then the remaining
explicit chiral symmetry breaking effects at order ${\cal O}(\alpha m_{res})$ are contained in the constants $\delta$ and $\delta_{m_{res}}$.
The value of $\delta_{m_{res}}$ in Tab.~\ref{tab:lec} should be compared with the value of $C_2$ determined from $\Delta m_{res}$.
In particular, one should have $\delta_{m_{res}}= 2 B_0 C_2$. One
sees that this combination is roughly two times too large. Since the LEC $\delta_{m_{res}}$ is found by linearly extrapolating to the chiral limit, it is possible that the discrepancy arises due to the omission of chiral logs. Recall that the logs are ${\cal O}(m_q \log m_q)$ compared to the linear terms and so could make a difference. Also note
that the value of $C_2$ is roughly an order of magnitude smaller than the
physical $\delta$ LEC. Because $\delta$ and $C_1$ have the same
charge dependence, it is their sum which is extracted from the fit. Thus, we must subtract off the amount $2 B_0 C_1$ from $\delta$ to obtain the physical LEC. This amounts to about a 20\% reduction from the fitted value of $\delta$. Note that the higher order effects in the residual mass also affect the logarithms in Eqs.~(\ref{eq:chpt mss}) and~(\ref{eq:chpt degenerate 1-loop 2}) which therefore do not cancel exactly in the meson mass-squared difference with QED switched on and off. However, these terms are order ${\cal O}(m\, \Delta m_{res})$ in the chiral expansion, so we neglect them.
}

{
The meson mass-squared splittings show little dependence on the sea quark mass. In Fig.~\ref{fig:sea quark dep} we show a representative sample of splittings for each sea quark mass. This is reasonable since we have not coupled the sea quarks to the photons, so the difference between ensembles is likely due to fluctuations in the gluon fields. In partially quenched chiral perturbation theory for 2+1 flavors~\cite{Bijnens:2006mk}, there exists a term at NLO that couples the valence quark charges to the sea quark masses. To account for this possibility, we include the $\delta_{sea}$ term in our fit (see Eq.~(\ref{eq:emcorrection}), and notice that only the charged mesons are affected). From Tab.~\ref{tab:lec} we see $\delta_{sea}$ is small, consistent with zero for the fit range including all masses, and only two standard deviations away from zero for the reduced range, which includes only the lightest two sea quark masses. The values of the other LEC's are largely unaffected by the $\delta_{sea}$ term, except $\delta$, which is reduced in the first case and increased in the second. Because the evidence is not strong in our data for such a term, we focus on the fits with $\delta_{sea}$ fixed to zero in the following.}

Using the results for the LEC's, we determine the quark masses as
follows.
{ As inputs, the three experimental values of
$m_{\pi^+}^2$, $m_{K^\pm}^2$, and $m_{K^0}^2$ are taken. We avoid the $\pi^0$ since we have not determined its true mass to NLO due to the lack of disconnected diagrams. We then iteratively solve the set of equations generated from Eq.~(\ref{eq:nlo pion mass}) for each meson for the three unknowns $m_u$, $m_d$, and $m_s$. At this stage, the quark masses are the
bare lattice values, including the shift due to the residual mass.
}

Using the non-perturbatively determined quark mass renormalization
constant $1/Z_m=Z_S$=0.62(4)~\cite{RBC NP Zm}, we obtain the values of the light quark masses,
$m^{\overline{\rm MS}}(2 {\rm GeV})=Z_m(m_q+m_{\rm res})$, and ratios  shown in Tab.~\ref{tab:quark masses}. The error on $Z_S$
reflects a statistical and a systematic error from the choice of $\Lambda_{QCD}$ in the range 250-300 MeV, which have been added in quadrature.  
As the bare quark mass range used in the fit to meson mass-squared is
reduced from { 0.0446 to 0.03, the physical quark masses change by about
one statistical standard deviation, or less. The only change outside of one standard deviation is for the quark mass ratios which are determined very precisely.
The ratio of up quark to down quark mass comes out to be close to
one-half while the strange quark to average up-down quark mass is about 28 to
29, depending on the fit range.}

{ Knowing the physical up and down quark masses, the physical charged-neutral
pion mass splitting is found from the LEC's in Tab.~\ref{tab:lec}.
Dividing the mass-squared difference by $m_{\pi^+}+m_{\pi^0}$, we find $m_{\pi^+}-m_{\pi^0}=3.89(17)$ MeV using all data, and 4.12(21) MeV using quark masses less than or equal to 0.03,
which is somewhat less than the physical value of 4.5936 MeV\cite{Yao:2006px}.
The part of this difference arising from the up-down quark mass
difference was estimated a long time ago to be about 0.17(3)
MeV~\cite{Gasser:1984gg} and more recently, 0.32 (20) MeV \cite{Bijnens:1996kk}.
In addition, there are still several systematic errors like non-zero
lattice spacing and finite volume effects to be addressed in the
calculation of the mass splitting, so the level of disagreement is not
surprising, perhaps even encouraging. It is interesting to note that almost the entire mass difference comes from the leading term ($\sim 98\%$) since the physical up and down quark masses are so small (the omission of disconnected diagrams for the $\pi^0$ mass has a very small effect here).
Similarly, we find the EM part of the kaon mass difference, 
$m_{K^+} - m_{K^0} = 1.443(55)$ MeV for the resricted range (1.441(42) MeV for the whole range).
QED effects make the charged kaon heavier, just as for the pions. In this case, the leading term also dominates, but the ${\cal O}(\alpha m_s)$ terms contribute about 23\%. Ultimately, the pure QCD $B_0 (m_u - m_d)$ term dominates the physical states and makes the neutral kaon heavier by 3.972(27) MeV\cite{Yao:2006px}.

{ From Eq.~(\ref{eq:dashen break}) the breaking of Dashen's theorem at ${\cal O}(\alpha m_q)$ is found to be $\Delta_{\rm EM} = 0.337(40)$ or 0.264(43), using the restricted fit range.
This is somewhat smaller than the large $N_c$ estimate reported in~\cite{Bijnens:1996kk}, $\Delta_{\rm EM} = 0.85(24)$. Following~\cite{Aubin:2004fs}, $\Delta_{\rm EM}$ would have to be roughly 10 in our calculation to render the up quark massless.
}
\subsection{systematic error estimate}
\label{subsec:error estimate}

We now turn to a discussion of the systematic errors in our calculation.
In the calculation of the neutral pion correlation function, we ignored
the contribution from disconnected diagrams.
{ Disconnected diagrams contribute to $m_{\pi^0}^2$ (and $m_\eta^2$) at
${\cal O}(\alpem m_q)$ \footnote{We have sought a proof that such contributions enter at ${\cal O}(\alpha m_q^2)$. In fact, one can show the leading contribution, where only a single photon is exchanged between the quark loops, vanishes. Likewise, any diagram where one photon and any even number of gluons are exchanged also vanishes if one considers each quark loop to be made from renormalized propagators and vertices only. 
However, if more general diagrams are considered, like the above but including a gluon exchange between two separated quark propagators from the same loop, the proof no longer holds. In any case, we expect these contributions to be suppressed, at least in the weak coupling limit of QCD.}
 so this is potentially a significant effect. However, we avoided using the $\pi^0$ mass to determine the quark masses, so this will not affect those estimates. And as mentioned above,
the physical $\pi^+-\pi^0$ mass splitting is dominated by the
LO contribution since the up and down quark masses are so small.
The main effect would be to alter the value of the LEC $\delta_0$ for the $\pi^0$ and $\eta$ mesons which we have assumed has the same value for all pseudoscalar mesons.
}
 
In any study of the EM interactions, finite volume effects may be
significant as the photons are massless (and unconfined).
In order to get a rough estimate for the size of this effect, we
examine the finite volume effect to $\delta$ using the
vector-saturation
model~\cite{Das:1967it,Bardeen:1988zw,Ecker:1988te,Donoghue:1993hj,Harada:2004qn}
as an example\footnote{ 
 More correctly, a parameter representing non-resonance contributions,
 which turns out to be small \cite{Ecker:1988te}, is set to zero.
}.
Assuming that the finite size effect purely due to QCD cancels in the
difference between $m_{\pi^+}$ and $m_{\pi^0}$, this model estimates the mass difference in a finite volume,
$\Delta_{\pi,\rm EM}=m_{\pi^+}^2-m_{\pi^0}^2$, to be 
\begin{equation} 
  \Delta_{\pi,\rm EM}(L) 
 = 
 \frac{3\,\alpha}{4\pi}\, 
 \frac{1}{a^2}\, 
 \frac{2^4\cdot \pi^2}{N} 
 \sum_{q \in \widetilde{\Gamma}^\prime} 
 \frac{(a m_\rho)^2 (a m_A)^2} 
      {\widehat{q}^{\,2} 
       \left(\widehat{q}^{\,2} + (a m_\rho)^2\right) 
       \left(\widehat{q}^{\,2} + (a m_A)^2\right)} \, , 
\end{equation} 
where $N$ is the total number of sites, $\widetilde{\Gamma}^\prime$ is the first Brillouin zone in the momentum space apart from $q_\mu = 0$, and
\begin{eqnarray} 
 \widehat{q}^{\,2} 
 &\equiv&  
 \sum_{\mu = 0}^3 \widehat{q}_\mu^{\,2} \, , 
  \nonumber \\ 
 \widehat{q}_\mu 
 &\equiv& 
 2\,\sin \left(\frac{a q_\mu}{2}\right) \, . 
\end{eqnarray} 
Applying our lattice volume and taking the ratio to that in the infinite
volume, we obtain
\begin{equation} 
 \frac{
        \Delta_{\pi,\rm EM}(\infty) 
        }
      {
        \Delta_{\pi,\rm EM}(L\approx1.9\ {\rm fm}) 
        } 
 = 1.10 \, .
  \label{eq:ratio_FV} 
\end{equation} 
Thus we find roughly a $+10$\% increase in $\delta$.
We expect a similar size of correction for the other
$\delta_i$'s.
The shift in $\delta$ could
affect the determination of the quark masses.
Shifting the value of physical $\delta-2 B_0 C_1$ by 10\%,
we find the quark masses change by less than 1\%. 
Thus we conclude that the finite volume effect on the quark masses due to the EM interaction is negligible.
However, removing this effect will enhance the pseudoscalar mass
splitting itself significantly.

We did not take into account the renormalization of the quark masses due to the EM interaction.
However, the QED part of the renormalization is expected to change the quark mass  by ${\cal O}(\alpem)\sim 1 \%$.
Since this is well within the statistical and other uncertainties
discussed already, we ignore this effect.
In future calculations it will be a simple matter to include these
effects directly in the non-perturbative renormalization calculation of
$Z_m$.

{
The use of the quenched approximation for the QED gauge fields 
results in a leading error of order $\alpem\,\alpha_S^2$ in correlation functions since the
dynamical quarks do not interact via the quenched photons.
However, as noted in~\cite{Bijnens:2006mk} the sea quark charge effects that enter at ${\cal O}(\alpha m_{sea})$ can be dealt with in two ways. First, the log terms come with known coefficients so can be subtracted from the lattice results before fitting for the LEC's. Second, the LEC's which come with sea quark dependence cancel out of some mass squared differences. The latter is operable here but not the former since we did not fit to the chiral logs as they are not known for the $N_f=2$ partially quenched case. We expect sea quark effects to be small in this present study since they were treated explicitly as neutral particles with respect to the EM interaction.
}

{
Finally, we note that changing our bare quark mass fit range from 
0.015-0.0446 (less than 1/2 $m_s$ to $m_s$) to 0.015-0.03 has
little effect on the LEC's or quark masses. Experience in the pure QCD case has shown that this range of quark masses is likely to be beyond the range of applicability of chiral perturbation theory~\cite{Aoki:2004ht}, and $\chi^2$/dof does decrease for the restricted range. So, being conservative,
we take as central values those results determined from the pseudoscalar
two-point function with a quark mass fit range
$0.015\le m_f \le 0.03$. There is some uncertainty introduced by the inclusion of the LEC $\delta_{sea}$,
which we set to zero or found to be small (or zero within errors) when
left as a free parameter. Since the evidence is uncertain for such a term in our data, and the sea quarks are not charged, we stick with the fit with $\delta_{sea}=0$ to quote our results for the quark masses.
Thus, from the third row of Tab.~\ref{tab:quark masses} our final
values for the quark masses are
\begin{eqnarray}
\label{eq:mu}
m^{\overline{\rm MS}}_u(2\, {\rm GeV}) &=& 3.02(27)(19) \mbox{ MeV},\\
m^{\overline{\rm MS}}_d(2\, {\rm GeV}) &=& 5.49(20)(34) \mbox{ MeV},\\
\label{eq:md}
m^{\overline{\rm MS}}_{ud}(2 \,{\rm GeV}) &=&4.25(23)(26) \mbox{ MeV},\\
\label{eq:mud}
m^{\overline{\rm MS}}_s(2 {\rm GeV})&= &119.5(56)(74) \mbox{ MeV},\\
\label{eq:ms}
m_u/m_{d} &=& 0.550(31),\\
\label{eq:mu/md}
m_s/m_{ud} &=& 28.10(38).
\label{eq:ms/mud}
\end{eqnarray}
The first error is statistical, the second from the error on the
renormalization constant $Z_m$.

Note that uncertainties due to the absence of the strange sea
quark and finite lattice spacing were not considered in the above
discussion of errors.}
We leave these issues for future work where they will be addressed explicitly by using $2+1$ flavor DWF gauge configurations at two lattice spacings.

\section{summary} 
\label{sec:summary} 

In this work, we have determined the electromagnetic splittings of the
pseudoscalar meson masses by calculating correlation functions in a
combined background of QCD+QED gauge fields which were, however,
generated separately.
The gluon configurations came from a recent two flavor domain wall
fermion simulation by the RBC Collaboration~\cite{Aoki:2004ht} while the
QED configurations were generated in the quenched approximation.

{ The highly correlated nature of the calculations allows very small effects from QED to be observed, even though the naive statistical errors on hadron masses are as large, or larger than the mass splittings themselves. This was seen in the original calculation~\cite{Duncan:1996xy} as well. Here we have gone a step further by averaging masses computed with $\pm$ electric charge on each configuration to cancel ${\cal O}(e)$ noise on each configuration.
This has lead to extremely small statistical errors on LEC's and 
physical ratios like $m_u/m_d$.}

The charged to neutral pion mass splitting was found to be
$m_{\pi^+}-m_{\pi^0}=4.12(21)$ MeV (for the restricted quark mass fit range) compared to the experimental value
4.5936 MeV\cite{Yao:2006px}, 0.17(3) or 0.32(20) MeV of which is due to the up-down
quark mass difference alone~\cite{Gasser:1984gg,Bijnens:1996kk}.
A simple model calculation leads to an estimate of $\sim 10$\% finite
volume effect.
Similarly, the kaon splitting arising from electromagnetism is 1.443(55) MeV (note that it is {positive}).
We emphasize that the calculations reported on here were carried out for
the physical value of $\alpem$, and we take these results to be
encouraging.
In the pioneering work by Duncan, Eichten, and
Thacker~\cite{Duncan:1996xy} which used the quenched approximation for
both QCD and QED and employed Wilson fermions at a single coarse lattice
spacing, the pion mass splitting was found to be 4.9(3) MeV.
Another more recent quenched calculation finds a value that is somewhat
higher still~\cite{Namekawa:2005dr}. 
 
{ Using the physical pseudoscalar meson masses as inputs, we were able to fix the values of the light
quark masses, $m_u$, $m_d$, and $m_s$, including effects of QED and violations of Dashen's theorem through ${\cal O}(\alpha m_q)$. These are given in Eqs.~(\ref{eq:mu})~-~(\ref{eq:ms/mud}).
}

In this work we have neglected the contributions to the neutral
masses arising from disconnected valence quark loop diagrams in the
two-point correlation functions because they are difficult to compute
precisely.
This leads to an uncertainty of
${\cal O}(\alpha^2,\,\alpha m_q,(m_u-m_d)^2)$ in the neutral pion
mass-squared  (see Eq.~(\ref{eq:axialWT-massless}) and the discussion in
Sec.~\ref{subsec:disconnected}).
The inclusion of these diagrams is left for future work.
But notice that this type of diagram is necessary only for the
calculation of the $\pi^0$ mass and not the $K^0$, and we did not use $\pi^0$ mass to determine the quark masses.

{ 
Finally, we also computed the vector meson mass splitting and found
it to be quite small, $\sim0.5$ MeV, essentially zero, considering systematic uncertainties in our calculation. This is an interesting topic that we will investigate further in future studies.}

The study presented here nicely sets the stage for future work that will
focus on the 2+1 flavor dynamical DWF configurations generated
by the RBC and UKQCD collaborations\cite{Allton:2007hx,RBCUK243}, and those that will soon be generated by the LHPC, RBC, and UKQCD collaborations.
Using smaller quark masses, larger lattices, and at least two lattice
spacings, the accuracy of the meson and quark mass splittings will
improve significantly.
Present and future work also includes the electromagnetic splittings of
the baryons~\cite{Doi:2006xh}.

\section*{Acknowledgments} 

M.~H. thanks the RIKEN BNL Research Center (RBRC) for kind hospitality
during his visit.
T.~B. thanks the theory group at KEK where part of this work was
accomplished for their support and generous hospitality.
T.~B. thanks Norman Christ for valuable discussions on gauge fixing.
  {T.I. is} grateful to Johan Bijnens for
  illuminating discussions about chiral perturbation
  theory formulas.
T.~D. is supported by Special Postdoctoral Research Program of RIKEN and by U.S. DOE grant DE-FG05-84ER40154.
The QCDOC supercomputer at the RBRC was used for the numerical
calculations in this work.  Some analysis of the results was performed
using the RIKEN Super Combined Cluster (RSCC).
This work was supported by the RBRC and the U.S. Department of Energy
under Outstanding Junior Investigator grant DE-FG02-92ER40716 (T.B.),
and in part by the Grant-in-Aid of the Ministry of Education
(No. 18034011, 18340075, 18740167).

\appendix
\section{Generation of non-compact $U(1)$ gauge field configurations}
\label{app:gauge fixing}

 Here we describe the way to generate 
configurations of $U(1)$ gauge fields on the lattice 
in the non-compact formulation. 
 In the non-compact $U(1)$ lattice gauge theory, 
the gauge potential $A_{{\rm em},\mu}(x)$ is treated 
as a basic dynamical variable and 
put on the mid-point of the link $(x,\,x + \widehat{\mu})$ 
of the hypercubic lattice 
with topology $T^4$ and with lattice spacing equal to $1$. 
 The purely gauge action is 
\begin{eqnarray} 
 S_{{\rm NC} {\it U(1)}}
 &=& \sum_x \sum_{\mu,\nu = 0}^3 \frac{1}{4 e^2}\,
     \bigg( \del_\mu A_{\rm em,\nu}(x) - \del_\nu A_{\rm em,\mu}(x)
     \bigg)^2 \,,
\label{eq:U1GaugeFieldAction}
\end{eqnarray} 
where $\del_\mu$ denotes the forward difference operator 
\begin{equation}
 \del_\mu f(x) \equiv f(x + \widehat{\mu}) - f(x)\, .
  \label{eq:forwardDiffrence}
\end{equation}
 The gauge potential $A_{{\rm em},\,\mu}(x)$ 
is assumed to obey the periodic boundary condition. 
 Then the gauge potential is expressed in momentum space as 
\begin{eqnarray}
 A_{{\rm em}\,,\mu}(x)
 &=& 
 \frac{1}{\sqrt{V}} \sum_{p\in\widetilde{\Gamma}}
  e^{i p \cdot \left(x + \frac{\widehat{\mu}}{2}\right)}
  \widetilde{A}_{\mu}(p) \, .
 \label{eq:FD_for_A}
\end{eqnarray} 
 Here $\displaystyle{V \equiv \prod_{\mu=0}^3 N_\mu}$ 
with $N_\mu$ the number of sites along the $\mu$-th direction, 
and $\widetilde{\Gamma}$ denotes the first Brillouin zone, 
\begin{eqnarray} 
 \widetilde{\Gamma} 
 &=& 
 \left\{ 
  \left.
   p_\mu = \frac{2\pi}{N_\mu}\,m_\mu
  \right| 
  m_\mu
  = - \left(\frac{N_\mu}{2} - 1\right) , \cdots, -1, 0, 1,
    \cdots, \frac{N_\mu}{2} \right\} \, .
 \label{eq:defOfFirstBrioullianZone}
\end{eqnarray} 
 In the decomposition (\ref{eq:FD_for_A}), 
not all of the modes are independent of each other 
because $A_{{\rm em},\,\mu}(x)$ is real-valued.
 Using the reflection operator $R$ in 
the first Brillouin zone $\widetilde{\Gamma}$ 
\begin{eqnarray} 
 R(p)_\mu 
 &=& 
 \left\{ 
  \begin{array}{cl} 
   -p_\mu & \mbox{if}\ p_\mu \ne \pi \\ 
   \pi & \mbox{if}\ p_\mu = \pi 
  \end{array} 
 \right. \, , 
 \label{eq:defOfReflectionFirstBrioullianZone} 
\end{eqnarray} 
the reality condition is expressed in momentum space as 
\begin{eqnarray} 
 \left( 
  e^{i \frac{R(p)_\mu}{2}}\,\widetilde{A}_\mu(R(p)) 
 \right) 
 = 
 \left( 
  e^{i \frac{p_\mu}{2}}\,\widetilde{A}_\mu(p) 
 \right)^* \, . 
\end{eqnarray} 
 In terms of these variables, 
the action (\ref{eq:U1GaugeFieldAction}) becomes 
\begin{eqnarray} 
 S_G 
 &=& 
 \frac{1}{2\,e^2} 
 \sum_{p \in \widetilde{\Gamma}} 
 \sum_{0 \le \mu < \nu \le 3} 
 \left| 
    \widehat{p}_\mu \widetilde{A}_{\nu}(p) 
  - \widehat{p}_\nu \widetilde{A}_{\mu}(p) 
 \right|^2 \, , 
  \label{eq:gaugeActionMomentumSpace} 
\end{eqnarray} 
where 
\begin{eqnarray} 
 && 
 \widehat{p}_\mu 
 \equiv 
 2\,\sin \left(\frac{p_\mu}{2}\right) \, . 
\end{eqnarray} 

 Let us introduce a single fermion field $\psi(x)$ with a unit charge 
in the system. 
 The following discussion applies even when 
various matter fields with different charges coexist 
in so far as the minimum charge is redefined to be unity 
and the system is invariant under the gauge transformation of the form 
\begin{eqnarray}
 A_{\rm em,\mu}(x) 
 &\mapsto& 
  A_{\rm em,\mu}^\prime(x) 
  = A_{\rm em,\mu}(x) + \del_\mu \Lambda(x)\, ,
\label{eq:gauge_transformation1} \\
 \psi(x)
 &\mapsto& 
  \psi^\prime(x) 
   = e^{i \Lambda(x)}\,\psi(x) \, ,
\label{eq:gauge_transformation2} \\
 \overline{\psi}(x)
 &\mapsto& \overline{\psi}^\prime(x)
 =         \overline{\psi}(x)\,e^{-i \Lambda(x)} \, .
 \label{eq:gauge_transformation3}
\end{eqnarray}
 We recall that from Eq.~(\ref{eq:gauge_transformation1}) 
the Wilson line constructed from $A_{{\rm em},\,\mu}(x)$ 
behaves as usual 
\begin{eqnarray}
 e^{- i A_{\rm em,\mu}^\prime(x)}
 &=& 
 e^{i \Lambda(x)}\,e^{- i A_{\rm em,\mu}(x)}\, 
  e^{- i \Lambda(x + \widehat{\mu})} \, . 
\end{eqnarray} 
 The matter fields are coupled to the gauge potential 
through $e^{i A_{{\rm em}\,\mu}(x)}$ 
as in the compact lattice formulation. 
 We also remark 
that $\Lambda(x)$ parameterizing gauge transformation 
is not necessarily periodic 
as long as the fields transformed 
via Eqs.~(\ref{eq:gauge_transformation1}), 
(\ref{eq:gauge_transformation2}) 
and (\ref{eq:gauge_transformation3}) 
continue to satisfy respective boundary conditions. 
 The only condition for $\Lambda(x)$ required 
from this consideration is 
\begin{eqnarray} 
 \Lambda(x + N_\mu \widehat{\mu})  
 &=& 
 \Lambda(x) + 2 \pi r_\mu \quad (r_\mu \in {\mathbb Z}) \, . 
 \label{eq:bc_for_Lambda} 
\end{eqnarray} 
 The quantization condition for $r_\mu$ 
arises from the presence of matter fields with nonzero electric charge. 
 In general $\Lambda(x)$ satisfying the condition (\ref{eq:bc_for_Lambda}) 
can be written in the form 
\begin{equation} 
 \Lambda(x) 
 = 
 \sum_{\mu = 0}^3 2\pi\,r_\mu\,\frac{x_\mu}{N_\mu} 
 + 
 \Lambda^{(0)}(x) \, , 
  \label{eq:Lambda_Lambda^0} 
\end{equation} 
where $\Lambda^{(0)}(x)$ satisfies 
the periodic boundary condition. 
 $\Lambda^{(0)}(x)$ thus has Fourier decomposition 
\begin{equation}
   \Lambda^{(0)}(x)
 = \frac{1}{\sqrt{V}} \sum_{p \in \widetilde{\Gamma}}
   e^{i p \cdot x}\,\widetilde{\Lambda}(p) \, .
 \label{eq:FD_Lambda(0)}
\end{equation} 
 From the reality condition 
$\widetilde{\Lambda}(R(p)) = \left(\widetilde{\Lambda}(p)\right)^*$, 
the constant mode $\widetilde{\Lambda}(0)$ is real. 
 Using Eqs.~(\ref{eq:FD_for_A}), (\ref{eq:FD_Lambda(0)}), 
the gauge transformation (\ref{eq:gauge_transformation1}) 
for the gauge potential becomes in the momentum space 
\begin{eqnarray} 
 \widetilde{A}_{\mu}(p)
 &\mapsto& 
 \widetilde{A}^\prime_{\mu}(p)
 = 
 \widetilde{A}_{\mu}(p)
  + 2\pi\,r_\mu\,\frac{\sqrt{V}}{N_\mu}\,\delta_{p,\,0}
  + i\,\widehat{p}_\mu\,\widetilde{\Lambda}(p) \, .
\end{eqnarray} 
 $\widetilde{\Lambda}(0)$ acts only on the matter fields 
through $e^{i \Lambda(0)}$, which ranges over a compact space. 
 We can thus leave it unfixed in the gauge fixing procedure 
for the calculation 
of the expectation values of operators. 

 The non-compact formulation needs 
the explicit fixing of the $U(1)$ gauge symmetry. 
 We employ the Coulomb gauge fixing condition 
\begin{eqnarray} 
 \sum_{j=1}^3 \drvstar{j} A_{{\rm em},\,j}(x) = 0 \, , 
  \label{eq:CoulombGauge}
\end{eqnarray} 
where $\drvstar{\mu}$ is the backward difference operator  
\begin{equation} 
 \drvstar{\mu} f(x) 
 \equiv 
 f(x) - f(x - \widehat{\mu}) \, . 
\end{equation} 
 In momentum space 
the condition (\ref{eq:CoulombGauge}) becomes 
\begin{equation}
 \sum_{j=1}^3
 \widehat{p}_j\,\widetilde{A}_{j}(p) = 0 \, .
  \label{eq:Coulomb_mom}
\end{equation}
 The condition (\ref{eq:CoulombGauge}) 
is not sufficient to fix all $U(1)$ redundancy; 
only the redundancy corresponding to the parameters 
$\{\Lambda(p_0,\,{\bf p})\}_{{\bf p} \ne {\bf 0}}$, 
where ${\bf p} \equiv (p_1,\,p_2,\,p_3)$,  
is eliminated by this condition. 
 Accordingly, for ${\bf p} \ne 0$, 
a component $\widetilde{A}_j(p)$ with $p_j \ne 0$ 
is determined by the other two spatial components 
\begin{equation}
 \widetilde{A}_{j}(p)
 = - \frac{1}{\widehat{p}_j}\sum_{k \ne j}
     \widehat{p}_k\,\widetilde{A}_{k}(p) \, ,
  \label{eq:A_j_solved}
\end{equation} 
irrespective of whether $p_0 \ne 0$ or $p_0 = 0$. 
 The residual gauge symmetry is generated by
$\{\Lambda(p_0,\,{\bf 0})\}_{p_0 \ne 0}$ 
(spatially uniform gauge transformation) and $r_\mu$.
 They act only on 
$\{\widetilde{A}_{\mu=0}(p_0,\,{\bf 0})\}_{p_0 \ne 0}$ 
and the constant modes $\widetilde{A}_\mu(0)$ 
of $A_{{\rm em},\,\mu}(x)$ respectively. 
 For $p_0 \ne 0$, 
a given $\widetilde{A}_{0}(p_0,\,{\bf 0})$ will be converted 
to a gauge configuration
$\widetilde{A}^\prime_{0}(p_0,\,{\bf 0}) = 0$ on the same orbit 
via $\widetilde{\Lambda}(p_0,\,{\bf 0})$ given by 
\begin{equation}
 \widetilde{\Lambda}(p_0,\,{\bf 0})
 = i\,\frac{1}{\widehat{p}_0}\,
 \widetilde{A}_{0}(p_0,\,{\bf 0}) \, .
\end{equation}
 In this way we can impose a condition 
\begin{eqnarray}
 \widetilde{A}_{\mu = 0}(p_0,\,{\bf 0}) = 0 
  \quad (p_0 \ne 0) \, .
\end{eqnarray} 
to eliminate $\widetilde{\Lambda}(p_0,\,{\bf 0})$. 
 The remained ones are $r_\mu \in {\mathbb Z}$. 
 These can be eliminated by imposing the following condition on 
the constant modes 
\begin{eqnarray} 
 0 \le \widetilde{A}_\mu(0) < 2\pi\,\frac{\sqrt{V}}{N_\mu} \, . 
  \label{eq:residualConstantModes} 
\end{eqnarray} 
 Thus we eliminated harmful gauge redundancy. 

 The degrees of freedom (\ref{eq:residualConstantModes}), 
the counterpart of Wilson loops in lower dimensions,  
cannot be gauged away. 
 However such degrees of freedom do not play vital roles  
for the dynamics in four-dimensional gauge theory 
with sufficiently large volume. 
 Thus, we fix $\widetilde{A}_\mu(0)$ 
to the constants $c_\mu$ in the range $(\ref{eq:residualConstantModes})$ 
as a boundary condition  
\begin{eqnarray} 
 \widetilde{A}_\mu(0) = c_\mu \, . 
\end{eqnarray} 

 Now we turn to the description of generations of gauge configurations. 
 It is sufficient to concentrate on 
generating configurations in momentum space 
since the Fourier transformation allows to convert them 
into the ones in coordinate space. 

 First we consider a mode with ${\bf p} \ne 0$. 
 Without loss of generality we can then assume that $p_3 \ne 0$.
 The independent integrated variables are $\widetilde{A}_{\mu}(p)$
($\mu = 0,\,1,\,2$) 
while $\widetilde{A}_{3}(p)$ is given by Eq.~(\ref{eq:A_j_solved}) with $j=3$.
 Inserting such $\widetilde{A}_{3}(p)$ into the corresponding part of
Eq.~(\ref{eq:gaugeActionMomentumSpace}), we get
\begin{eqnarray}
 && \sum_{p_3 \ne 0} \frac{1}{2\,e^2}  
    \left[ \widehat{{\bf p}}^2 
     \left| \widetilde{A}_{0}(p) \right|^2 
    \right. 
 \nonumber \\ 
 && \qquad \qquad 
    \left. + \widehat{p}^{\,2}
     \sum_{k=1}^2 \left|\widetilde{A}_{k}(p)\right|^2 
    + \frac{\widehat{p}^{\,2}}{(\widehat{p}_3)^2}\sum_{j,\,k = 1}^2 
      \widehat{p}_j \widehat{p}_k\,
   {\rm Re}\,
   \left(
    \widetilde{A}_{j}(p)\,\left(\widetilde{A}_{k}(p)\right)^*
   \right)
 \right] \, ,
  \label{eq:action_for_p3_nonzero}
\end{eqnarray}
where 
\begin{eqnarray}
 \widehat{p}^{\,2} 
 &\equiv& 
 \sum_{\mu = 0}^3 (\widehat{p}_\mu)^2 \, , 
  \nonumber \\ 
 \widehat{{\bf p}}^2 
 &\equiv& 
 \sum_{j = 1}^3 (\widehat{p}_j)^2 \, . 
\end{eqnarray} 
 The action (\ref{eq:action_for_p3_nonzero}) 
gives the Boltzmann weight 
in generating configurations for the modes 
with ${\bf p}\ne {\bf 0}$. 
 However that form is not useful yet for generating configurations 
because the two spatial components are mixed with each other 
due to the terms 
\begin{equation} 
 \left( 
  \begin{array}{cc} 
   \left(\widetilde{A}_{1}(p)\right)^*,\, & 
   \left(\widetilde{A}_{2}(p)\right)^* 
  \end{array} 
 \right) 
 M 
 \left( 
  \begin{array}{c} 
   \widetilde{A}_{1}(p) \\
   \widetilde{A}_{2}(p) 
  \end{array} 
 \right) \, , 
\end{equation}  
where a $2\times 2$ real symmetric matrix $M$ takes the form 
\begin{equation} 
 M 
 = 
 \widehat{p}^{\,2} 
 \left( 
  \begin{array}{cc} 
   \displaystyle{ 
    1 + \frac{(\widehat{p}_1)^2}{(\widehat{p}_3)^2} 
   } 
   & 
   \displaystyle{ 
    \frac{\widehat{p}_1 \widehat{p}_2}{(\widehat{p}_3)^2} 
   } \\ 
   \displaystyle{ 
    \frac{\widehat{p}_1 \widehat{p}_2}{(\widehat{p}_3)^2} 
   } 
   & 
   \displaystyle{ 
    1 + \frac{(\widehat{p}_2)^2}{(\widehat{p}_3)^2} 
   } 
  \end{array} 
 \right) \, . 
\end{equation} 
 As this is the eigenvalue problem in two dimension, 
it is possible to resolve this mixing analytically. 
 This is an advantage of Coulomb gauge over the covariant Lorentz gauge. 
 The eigenvalues of $M$ are 
\begin{eqnarray} 
 m_- 
 &\equiv& 
 \widehat{p}^{\,2} \, , 
  \nonumber \\ 
 m_+ 
 &\equiv& 
 \widehat{p}^{\,2} 
 \left( 
    1 
  + \frac{(\widehat{p}_1)^2 + (\widehat{p}_2)^2} 
         {(\widehat{p}_3)^2} 
 \right) 
 \, . 
\end{eqnarray} 
 Those two eigenvalues become degenerate if and only if 
$\widehat{p}_1 = 0 = \widehat{p}_2$. 
 In this case, we can use the basis 
$(\widetilde{A}_{1}(p),\,\widetilde{A}_{2}(p))$ to generate 
a configuration according to the action 
\begin{equation} 
 \frac{h_p}{2\,e^2}\, 
 \widehat{p}^{\,2} 
 \left( 
  \left|\widetilde{A}_{1}(p)\right|^2 
  + 
  \left|\widetilde{A}_{2}(p)\right|^2 
 \right) 
 \quad 
 (\widehat{p}_1 = 0 = \widehat{p}_2)\, . 
\end{equation} 
 Here 
$h_p = 1$ for $R(p) = p$ and $h_p = 2$ for $R(p) \ne p$, 
because the action gets doubled due to 
the contribution from the complex conjugate partner 
in the latter case. 
 In the case 
that $(\widehat{p}_1)^2 + (\widehat{p}_2)^2 \ne 0$, 
$M$ can be diagonalized as 
\begin{equation} 
 M 
 = 
 O\,
 {\rm diag}\, 
 \left(m_-,\,m_+ \right)\,
 O^{-1} \, .
\end{equation} 
 Here the orthogonal matrix $O$ is 
\begin{equation} 
 O 
 = 
 \left( 
  \begin{array}{cc} 
    r_2 & r_1 \\ 
   -r_1 & r_2 
  \end{array} 
 \right) 
  \, , 
\end{equation} 
with 
\begin{equation} 
 r_j 
 \equiv 
 \frac{\widehat{p}_j} 
      {\sqrt{(\widehat{p}_1)^2 + (\widehat{p}_2)^2}} 
 \quad 
 (j = 1,\, 2) \, . 
\end{equation} 
 We first generate a configuration for 
$\left(\widetilde{A}_{-}(p),\,\widetilde{A}_{+}(p)\right)$ 
according to the action 
\begin{equation} 
 \frac{h_p}{2\,e^2} 
 \left( 
    m_- \left|\widetilde{A}_{-}(p)\right|^2 
  + m_+ \left|\widetilde{A}_{+}(p)\right|^2 
 \right) 
 \, . 
\end{equation} 
 A configuration for 
$\left(\widetilde{A}_{1}(p),\,\widetilde{A}_{2}(p)\right)$ 
will be obtained by using the rotation matrix $O$ as 
\begin{equation} 
 \left( 
  \begin{array}{c} 
   \widetilde{A}_{1}(p) \\ 
   \widetilde{A}_{2}(p) 
  \end{array} 
 \right) 
 = 
 O 
 \left( 
  \begin{array}{c} 
   \widetilde{A}_{-}(p) \\ 
   \widetilde{A}_{+}(p) 
  \end{array} 
 \right) \, . 
\end{equation} 

 For a mode with ${\bf p} = {\bf 0}$ but $p_0 \ne 0$, 
the temporal component 
$\widetilde{A}_{0}(p_0,\,{\bf 0})$ is zero. 
 The three spatial components are then integrated independently 
according to the action 
\begin{eqnarray} 
 \frac{h_p}{2\,e^2} 
 \left(\widehat{p}_0\right)^2 
 \sum_{j=1}^3 
 \left|\widetilde{A}_{j}(p)\right|^2 \, . 
\end{eqnarray}

\clearpage
\begin{table}
 \centering
 \begin{tabular}{ccc}
  \hline
 $m_{sea}$ & trajectories & configurations \\\hline
 0.02 & 706-5356 & 95\\
 0.03 & 695-6195 & 111\\
 0.04 & 605-5605\footnote{configurations 1805, 1855, 1905, and 1955 were excluded from our analysis to avoid possible effects of a  hardware failure as noted in~\cite{Aoki:2004ht}.} & 97\\
   \hline
 \end{tabular}
 \caption{QCD gauge configuration ensemble from~\cite{Aoki:2004ht} used for the measurements in this work. Configurations used for measurements are separated by 50 monte-carlo trajectories. Parameters for the ensembles are listed in Tab.~\ref{tab:qcd parameters}.}
 \label{tab:ensembles}
\end{table}
\begin{table}
 \centering
 \begin{tabular}{c|c}
  \hline
  $\beta_{\tiny\rm QCD}$ & 0.8 \\
  Lattice size      & $V=16^3\times 32$,\ \ $L_s$=12 \\
  domain-wall height& $M_5$=1.8\\
  Sea quark masses  & 0.02,\ 0.03,\ 0.04 \\
  $1/a$             & $1.691(53)$ GeV from $m_\rho$=770 MeV\\
                    & $1.688(21)(^{+69}_{-04})$ GeV from $r_0$=0.5 fm\\
  \hline
 \end{tabular}
 \caption{Simulation parameters in the generation of two-flavor
 dynamical QCD configurations, and the lattice spacing obtained in the
 pure QCD simulation. For details, see Ref.~\cite{Aoki:2004ht}.} 
 \label{tab:qcd parameters}
\end{table}
\begin{table}
 \centering
 \begin{tabular}{c|c}
  \hline
  $\beta_{\tiny\rm em}=2/e^2$ & 2 \\
  $e$ & $\sqrt{4\pi/137}$(physical value), 0.6, 1.0 \\
  domain-wall height& $M_5$=1.8\\
    $1/a$             & $1.659(37)$ GeV from $m_\rho$=775.8 MeV\\
  \hline
 \end{tabular}
 \caption{Simulation parameters in this work.
 The same value of $M_5$ as the sea quarks is used for the valence
 quarks.}
 \label{tab:qed parameters}
\end{table}
\begin{table}
 \centering
 \begin{tabular}{ccccccc}
  \hline
   $m_u$ &\multicolumn{1}{c}{$m_d=0.015$}&0.02&0.025&0.03&0.035&0.04  \\
  \hline
\hline
\multicolumn{7}{c}{$m_{sea}=0.02$} \\
\hline
0.015 & 0.2580(27) & 0.2759(25) & 0.2928(24) & 0.3090(23) & 0.3244(23) & 0.3392(22) \\
0.02 & 0.2760(25) & 0.2927(24) & 0.3087(23) & 0.3240(22) & 0.3388(21) & 0.3530(21) \\
0.025 & 0.2930(24) & 0.3088(23) & 0.3240(22) & 0.3386(21) & 0.3528(20) & 0.3665(20) \\
0.03 & 0.3092(23) & 0.3242(22) & 0.3387(21) & 0.3528(20) & 0.3664(20) & 0.3797(19) \\
0.035 & 0.3247(23) & 0.3390(21) & 0.3530(20) & 0.3665(20) & 0.3797(19) & 0.3926(19) \\
0.04 & 0.3396(22) & 0.3534(21) & 0.3668(20) & 0.3798(19) & 0.3926(19) & 0.4052(18) \\
\hline
\multicolumn{7}{c}{$m_{sea}=0.03$} \\
\hline
0.015 & 0.2633(22) & 0.2806(21) & 0.2972(20) & 0.3130(19) & 0.3281(19) & 0.3427(19) \\
0.02 & 0.2808(21) & 0.2970(20) & 0.3127(19) & 0.3277(18) & 0.3422(18) & 0.3563(17) \\
0.025 & 0.2974(20) & 0.3127(19) & 0.3276(18) & 0.3421(18) & 0.3561(17) & 0.3696(17) \\
0.03 & 0.3132(19) & 0.3279(18) & 0.3422(18) & 0.3560(17) & 0.3696(17) & 0.3827(16) \\
0.035 & 0.3285(19) & 0.3425(18) & 0.3562(17) & 0.3696(17) & 0.3827(16) & 0.3955(16) \\
0.04 & 0.3431(19) & 0.3566(17) & 0.3699(17) & 0.3829(16) & 0.3956(16) & 0.4080(15) \\
\hline
\multicolumn{7}{c}{$m_{sea}=0.04$} \\
\hline
0.015 & 0.2659(25) & 0.2835(24) & 0.3002(23) & 0.3162(23) & 0.3315(22) & 0.3463(22) \\
0.02 & 0.2836(24) & 0.3002(23) & 0.3160(23) & 0.3312(22) & 0.3459(22) & 0.3601(21) \\
0.025 & 0.3004(23) & 0.3161(23) & 0.3312(22) & 0.3458(21) & 0.3599(21) & 0.3736(21) \\
0.03 & 0.3165(23) & 0.3314(22) & 0.3459(21) & 0.3599(21) & 0.3736(21) & 0.3868(20) \\
0.035 & 0.3319(23) & 0.3462(22) & 0.3601(21) & 0.3737(21) & 0.3868(20) & 0.3997(20) \\
0.04 & 0.3467(22) & 0.3604(21) & 0.3739(21) & 0.3870(20) & 0.3998(20) & 0.4122(19) \\
  \hline
 \end{tabular}
 \caption{Summary of $u\bar d$ pseudoscalar meson masses obtained from fits to the pseudoscalar two-point correlation functions. Fit range is $9\le t \le16$ in
 each case. $\alpha =1/137$.}
 \label{tab:ps masses}
\end{table}

\begin{table}
 \centering
 \begin{tabular}{ccccccc}
  \hline  
  $m_{sea}$ &  \multicolumn{1}{c}{$m_u=0.015$}&0.02&0.025&0.03&0.035&0.04 \\
  \hline
0.02 &0.2575(27) & 0.2923(24) & 0.3236(22) & 0.3524(20) & 0.3794(19) & 0.4049(18) \\
0.03 & 0.2628(22) & 0.2966(20) & 0.3273(18) & 0.3558(17) & 0.3825(16) & 0.4078(15) \\
0.04 & 0.2654(25) & 0.2997(23) & 0.3308(22) & 0.3596(21) & 0.3865(20) & 0.4119(19) \\
  \hline
 \end{tabular}
 \caption{Summary of neutral $u\bar u$ pseudoscalar meson masses obtained from fits to the pseudoscalar two-point correlation functions.
 Fit range is $9\le t \le16$ in
 each case. $\alpha =1/137$.}
 \label{tab:uu masses}
\end{table}

\begin{table}
 \centering
 \begin{tabular}{ccccccc}
  \hline  
  $m_{sea}$ & \multicolumn{1}{c}{$m_d=0.015$}&0.02&0.025&0.03&0.035&0.04 \\
  \hline
0.02 & 0.2566(27) & 0.2913(24) & 0.3226(22) & 0.3514(20) & 0.3783(19) & 0.4037(18) \\
0.03 &0.2620(22)& 0.2957(20) & 0.3263(18) & 0.3547(17) & 0.3814(16) & 0.4066(15) \\
0.04 & 0.2645(25) & 0.2988(23) & 0.3298(22) & 0.3585(21) & 0.3854(20) & 0.4108(19) \\
  \hline
 \end{tabular}
 \caption{Summary of neutral $d\bar d$ pseudoscalar meson masses obtained from fits to the pseudoscalar two-point correlation functions.
 Fit range is $9\le t \le16$ in
 each case. $\alpha =1/137$.}
 \label{tab:dd masses}
\end{table}

\begin{table}
 \centering
 \begin{tabular}{ccccccc}
  \hline  
  $m_{sea}$ & \multicolumn{1}{c}{$m_u=0.015$}&0.02&0.025&0.03&0.035&0.04 \\
  \hline
0.02 & 0.3524(22) & 0.3658(20) & 0.3788(20) & 0.3916(19) & 0.4042(18) & 0.4164(18) \\
0.03 & 0.3557(18) & 0.3688(17) & 0.3818(16) & 0.3945(16) & 0.4070(15) & 0.4192(15) \\
0.04 & 0.3594(22) & 0.3728(21) & 0.3859(20) & 0.3987(20) & 0.4112(19) & 0.4235(19) \\
  \hline
 \end{tabular}
 \caption{Summary of charged $u\bar s$ pseudoscalar meson masses obtained from fits to the pseudoscalar two-point correlation functions.
 $m_s=0.0446$. Fit range is $9\le t \le16$ in
 each case. $\alpha =1/137$.}
 \label{tab:us masses}
\end{table}

\begin{table}
 \centering
 \begin{tabular}{cccccccc}
  \hline  
  $m_{sea}$ & \multicolumn{1}{c}{$m_d=0.015$}&0.02&0.025&0.03&0.035&0.04& 0.0446 \\
  \hline
0.02 & 0.3512(22) & 0.3645(20) & 0.3775(20) & 0.3903(19) & 0.4028(18) & 0.4150(18) & 0.4261(17)\\
0.03 & 0.3546(18) & 0.3677(17) & 0.3806(16) & 0.3933(16) & 0.4057(15) & 0.4179(15) & 0.4289(15)\\
0.04 & 0.3582(22) & 0.3715(21) & 0.3846(20) & 0.3973(20) & 0.4098(19) & 0.4220(19)  &0.4330(19)\\
  \hline
 \end{tabular}
 \caption{Summary of neutral $d\bar s$ and $s\bar s$ pseudoscalar meson masses obtained from fits to the pseudoscalar two-point correlation functions. $m_s=0.0446$.
 Fit range is $9\le t \le16$ in
 each case. $\alpha =1/137$.}
 \label{tab:ds masses}
\end{table}
\begin{table}
 \centering
 \begin{tabular}{cccccccc}
  \hline
 meson& $\msea$ & \multicolumn{1}{c}{$\mval=0.015$}&0.02&0.025&0.03&0.035&0.04  \\
  \hline
ud & 0.02 &  0.5305(60) &  0.5449(48) &  0.5586(40) &  0.5724(35) &  0.5865(31) &  0.6010(29)  \\
ud & 0.03 &  0.5494(79) &  0.5614(63) &  0.5741(53) &  0.5873(47) &  0.6010(42) &  0.6151(39)  \\
ud & 0.04 &  0.5695(132) &  0.5809(104) &  0.5930(87) &  0.6059(74) &  0.6193(65) &  0.6330(58)  \\
uu & 0.02 & 0.5304(60) & 0.5449(47) & 0.5586(40) & 0.5725(35) & 0.5867(31) & 0.6012(29)  \\
uu & 0.03 & 0.5498(78) & 0.5617(62) & 0.5743(53) & 0.5876(47) & 0.6013(42) & 0.6155(39)  \\
uu & 0.04 & 0.5691(130) & 0.5807(104) & 0.5930(86) & 0.6060(74) & 0.6195(65) & 0.6333(57)  \\
dd & 0.02 & 0.5300(60) & 0.5444(48) & 0.5580(40) & 0.5717(35) & 0.5858(31) & 0.6003(29)  \\
dd & 0.03 & 0.5493(79) & 0.5612(63) & 0.5737(53) & 0.5868(47) & 0.6005(42) & 0.6145(39)  \\
dd & 0.04 & 0.5691(132) & 0.5804(105) & 0.5925(87) & 0.6053(74) & 0.6187(65) & 0.6324(58)  \\
  \hline
 \end{tabular}
 \caption{Summary of degenerate vector meson masses obtained from fits to the vector two-point correlation functions, averaged over polarizations. Fit ranges are $5\le t \le16$, $6\le t \le16$, and $7\le t \le16$, for $\msea=0.02, 0.03$, and 0.04, respectively. $\alpha =1/137$.}
 \label{tab:vec masses}
\end{table}
\begin{table}
 \centering
 \begin{tabular}{ccc|ccc}
  \hline
  $a_v$ & $b_v$ & $\chi^2$/dof & $c_v$ & $d_v$ & $\chi^2$/dof \\
  \hline
   0.461(11)   &  2.04(17)   & 0.20(89) &
   0.00028(20) & -0.0015(31) & 2.9(34)\\
  \hline
 \end{tabular}
 \caption{The results of the chiral extrapolation of $m_{\rho^\pm}$ and
 $m_{\rho^\pm}-m_{\rho^0}$.}
 \label{tab:vector}
\end{table}
\begin{table}
 \centering
 \begin{tabular}{cccccccc}
  \hline
  fit range & $\delta$ &$\delta_0$ &$\delta_+$ &$\delta_-$ &$\delta_{sea}$ & $\delta_{m_{res}}$ & $\chi^2$/dof\\
  \hline
  0.015-0.0446 & 4.62 (18) $\times 10^{-4}$ & 0.0080 (12) & 0.01129 (24) & 0.01746(33) & - & 6.8 (10) $\times 10^{-5}$ &1.7(1.2)\\
     & 4.45 (56) $\times 10^{-4}$ & 0.0080 (12) & 0.01132 (23) & 0.01741(29) & $2.5(8.4)\times10^{-4}$ & 6.8 (10) $\times 10^{-5}$ &1.7(1.2)\\
0.015-0.03& 4.85 (21) $\times 10^{-4}$ & 0.0077 (20) & 0.01059 (32) & 0.01696(40) & - & 7.9 (14) $\times 10^{-5}$ &1.4(1.4)\\  
& 6.46 (86) $\times 10^{-4}$ & 0.0077 (20) & 0.01048 (32) & 0.01701(40) & -0.0028 (15) & 7.9 (14) $\times 10^{-5}$ &0.14(25)\\  
  \hline
 \end{tabular}
 \caption{The results of the NLO fit to the meson
 mass squared differences. The first column refers to the quark mass range, both sea and valence, used in the fit.}
 \label{tab:lec}
\end{table}
\begin{table}
 \centering
 \begin{tabular}{ccccc}
  \hline
  fit range & $B_0$ & $L_5-2L_8$ & $L_4-2L_6$ & $\chi^2$/dof\\
  \hline
  0.015-0.0446 & 2.172(67) & 5.7 (36) $\times 10^{-5}$ & -0.99 (28) $\times 10^{-4}$ & 0.91(30)\\
  0.015-0.03 & 2.14(13) & -8.6 (53) $\times 10^{-5}$ & -6.4 (77) $\times 10^{-5}$ & 0.08(6)\\
  \hline
 \end{tabular}
 \caption{The results of the $\alpha=0$ NLO fit to the meson
 masses. The first column refers to the quark mass range, both sea and valence, used in the fit.}
 \label{tab:alpha=0 lec}
\end{table}
\begin{table}
 \centering
 \begin{tabular}{ccccccc}
  \hline
fit range & $m_u$ & $m_d$ & $m_{ud}$ & $m_s$ & $m_u/ m_d$
&$m_s/m_{ud}$ \\
\hline
0.015-0.0446 & 2.96 (13) & 5.47 (11) & 4.21 (12) & 122.5 (27) & 0.540(14) & 29.08(30)\\
0.015-0.03 & 3.02 (27) & 5.49 (20) & 4.25 (23) & 119.5 (56) & 0.550(31) & 28.10(38)\\
\hline
 \end{tabular}
 \caption{Light quark masses evaluated for physical meson masses. All values are in MeV and given in the $\overline{MS}$ scheme at renormalization scale $\mu=2$ GeV. The first column (``fit range") refers to the range of bare quark masses used in the meson mass-squared fit. $\alpha=1/137$.}
 \label{tab:quark masses}
\end{table}
\clearpage

\begin{figure}
 \centering
   \includegraphics*[width=0.7 \textwidth,clip=true]
   {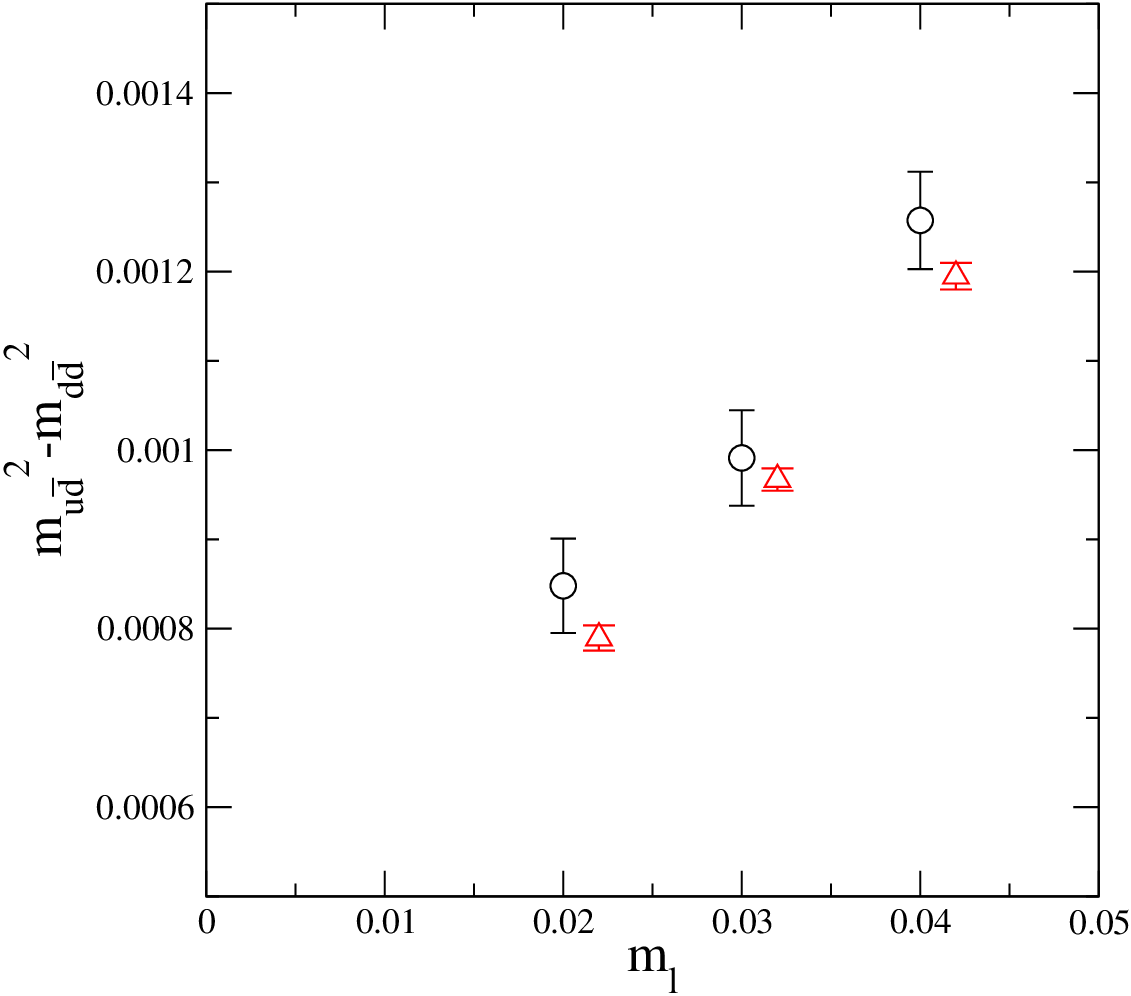}
 \caption{One of the pseudoscalar mass-squared splittings computed
 for $e=1$ (circles) and also averaged over $e=\pm1$ (triangles). The latter has dramatically reduced statistical error (and is shifted for clarity).}
 \label{fig:e+-1 trick}
\end{figure}

\begin{figure}
 \centering
   \includegraphics*[width=0.7 \textwidth,clip=true]
   {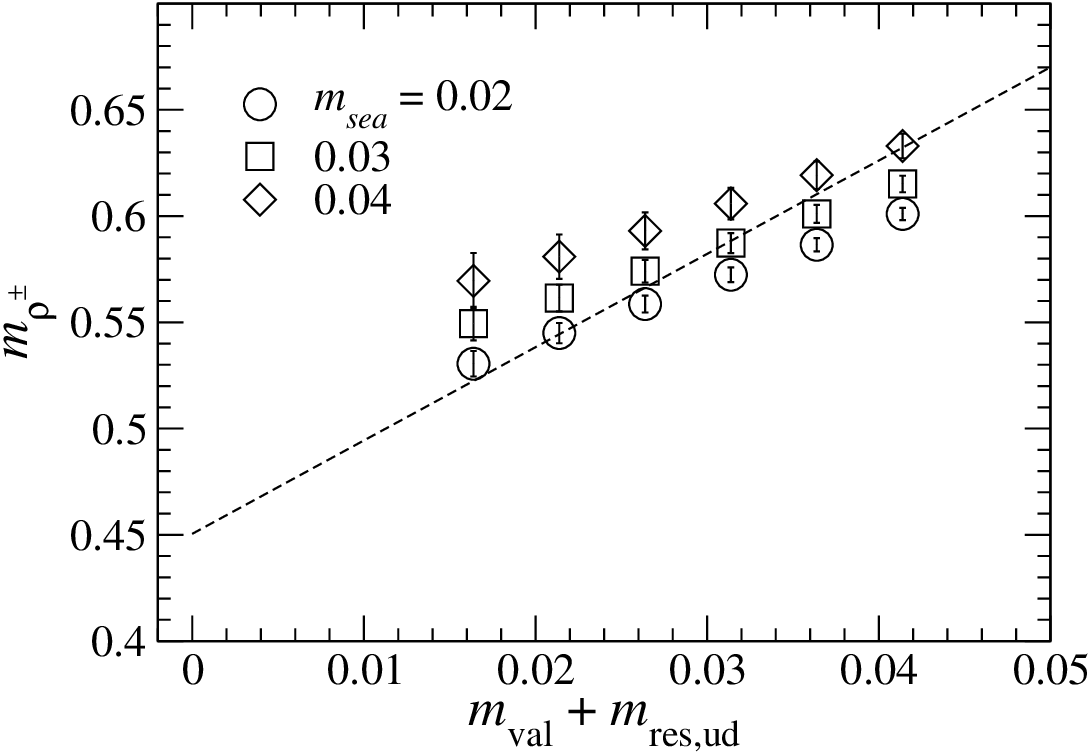}
 \caption{The quark mass dependence of charged vector meson mass and its
 chiral extrapolation using the $m_{sea}=m_{val}$ data and Eq.~(\ref{eq:vec1}).}
 \label{fig:vec_mass}
\end{figure}
\begin{figure}
 \centering
   \includegraphics*[width=0.7 \textwidth,clip=true]
   {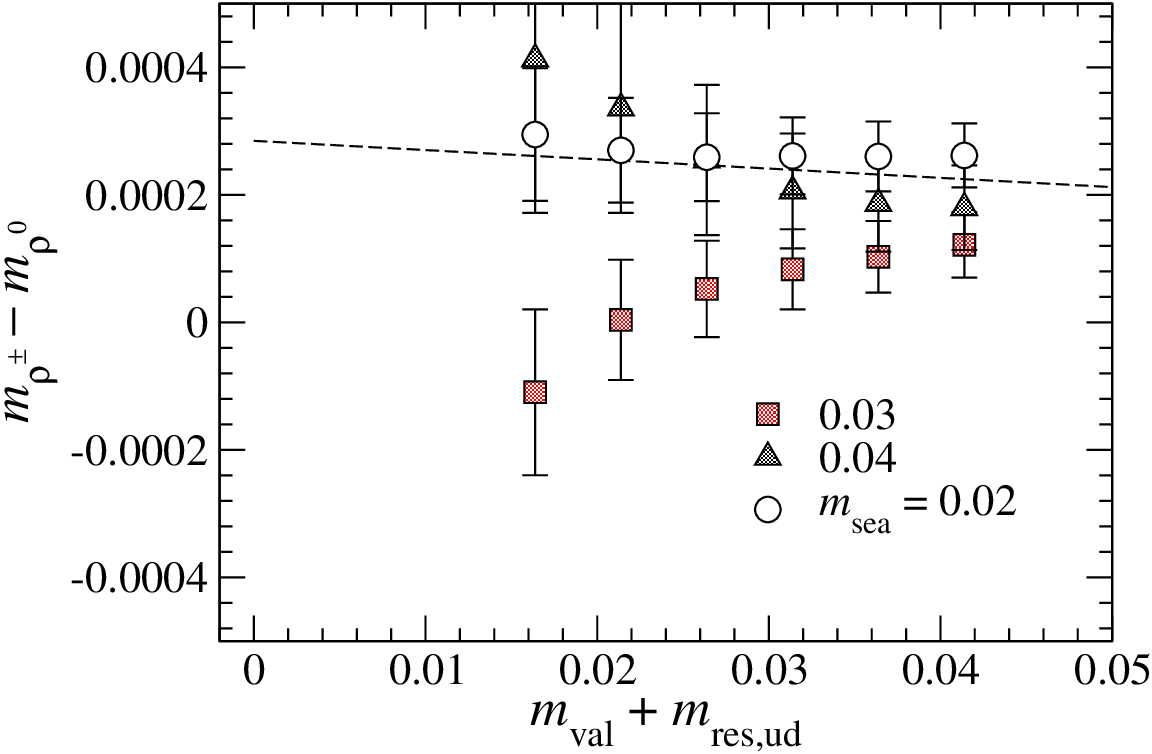}
 \caption{The quark mass dependence of the $\rho^{+}$-$\rho^0$
 mass splitting. The dashed line is a fit to Eq.~(\ref{eq:vec2}).}
 \label{fig:vec_massdiff}
\end{figure}
\begin{figure}
 \centering
   \includegraphics*[width=0.7 \textwidth,clip=true]
   {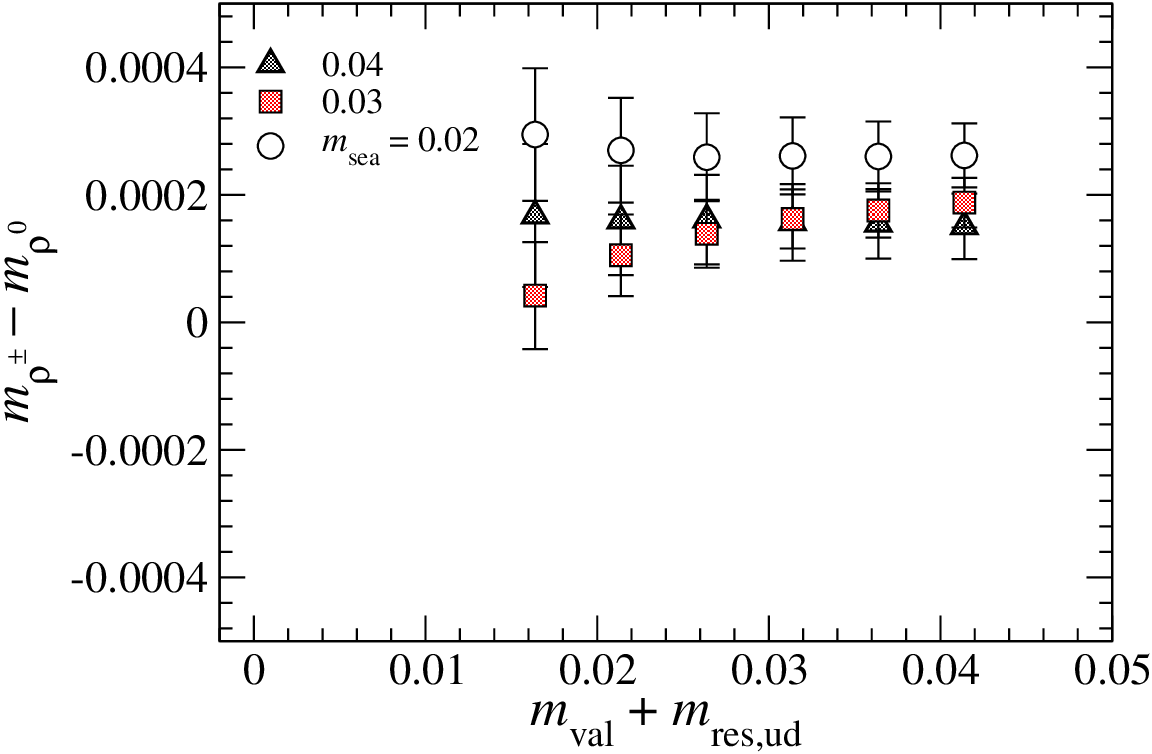}
 \caption{Same as Fig.~\ref{fig:vec_massdiff} except the minimum
 distance in the mass fits is $t=5$ for all cases.}
 \label{fig:vec_massdiff tmin5}
\end{figure}

\begin{figure}
 \centering
 \includegraphics*[width=0.6 \textwidth,clip=true]
 {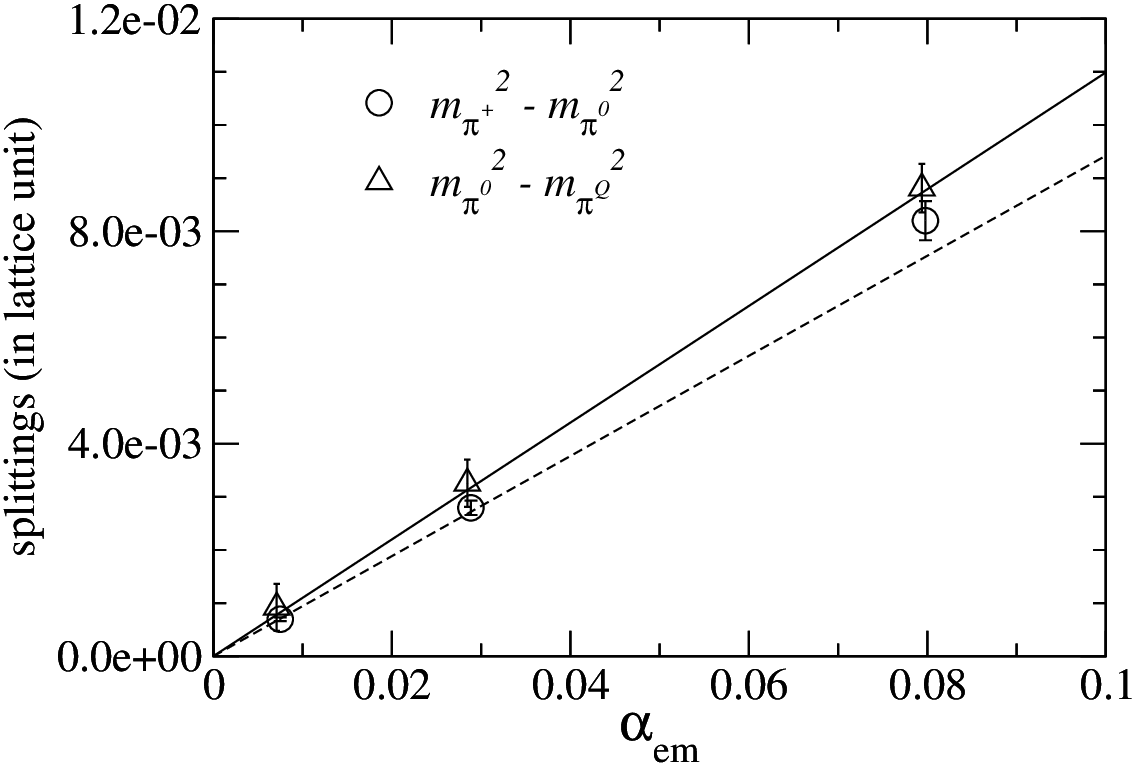}
 \caption{The $\alpha_{\rm em}$ dependence of the splittings.
 The results from $\la P\,P\ra$ are shown.}
 \label{fig:e-dep of splittings}
\end{figure}
\begin{figure}
 \centering
 \includegraphics*[width=0.6 \textwidth,clip=true]
 {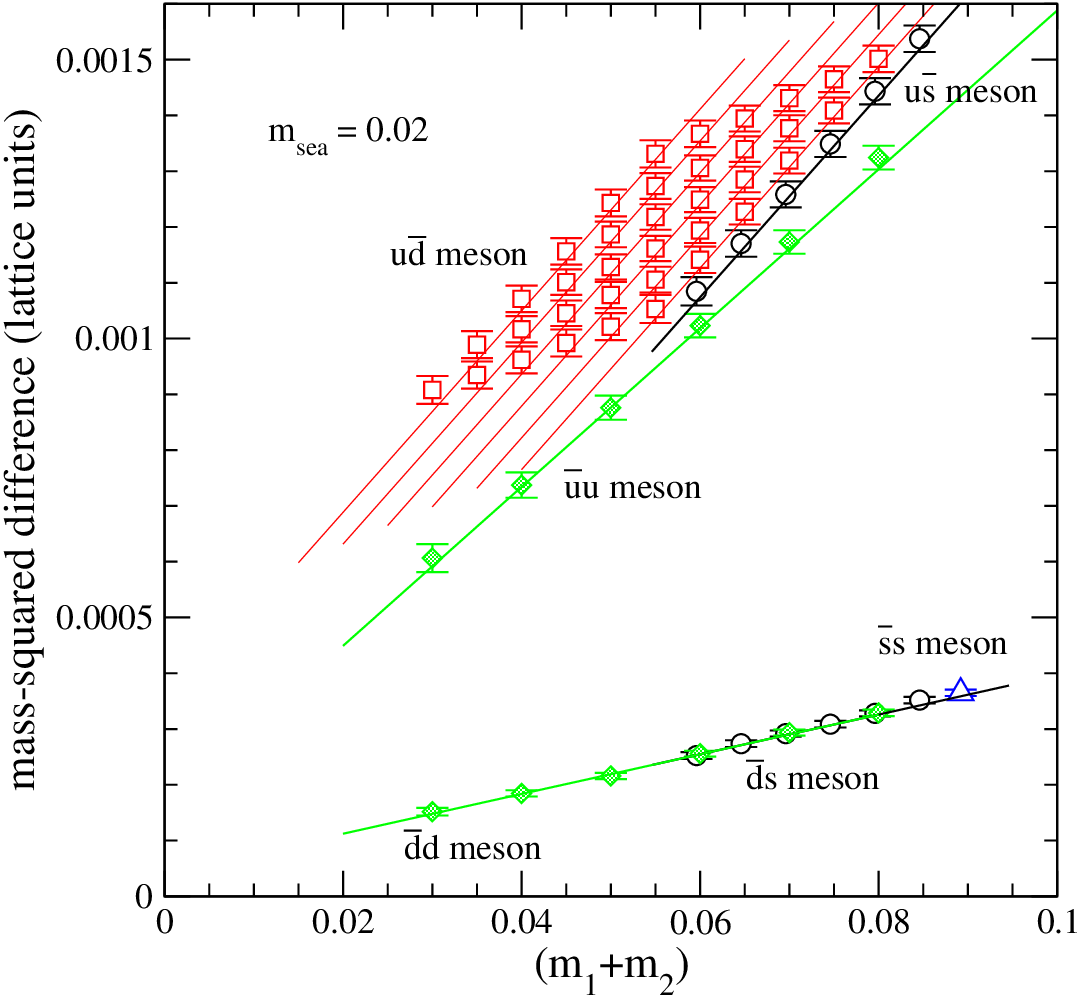}
 \caption{The pseudoscalar mass-squared splitting for $m_{sea}=0.02$. Each point corresponds to the mass-squared computed with $\alpha=1/137$ minus the same quantity computed with $\alpha =0$. The labels in the
 figure correspond to the charges of the quarks, $i.e$, $u\bar d$ means $Q_u=2/3$ and $Q_{\bar d} = 1/3$.}
 \label{fig:ps splittings}
\end{figure}
\begin{figure}
 \centering
 \includegraphics*[width=0.6 \textwidth,clip=true]
 {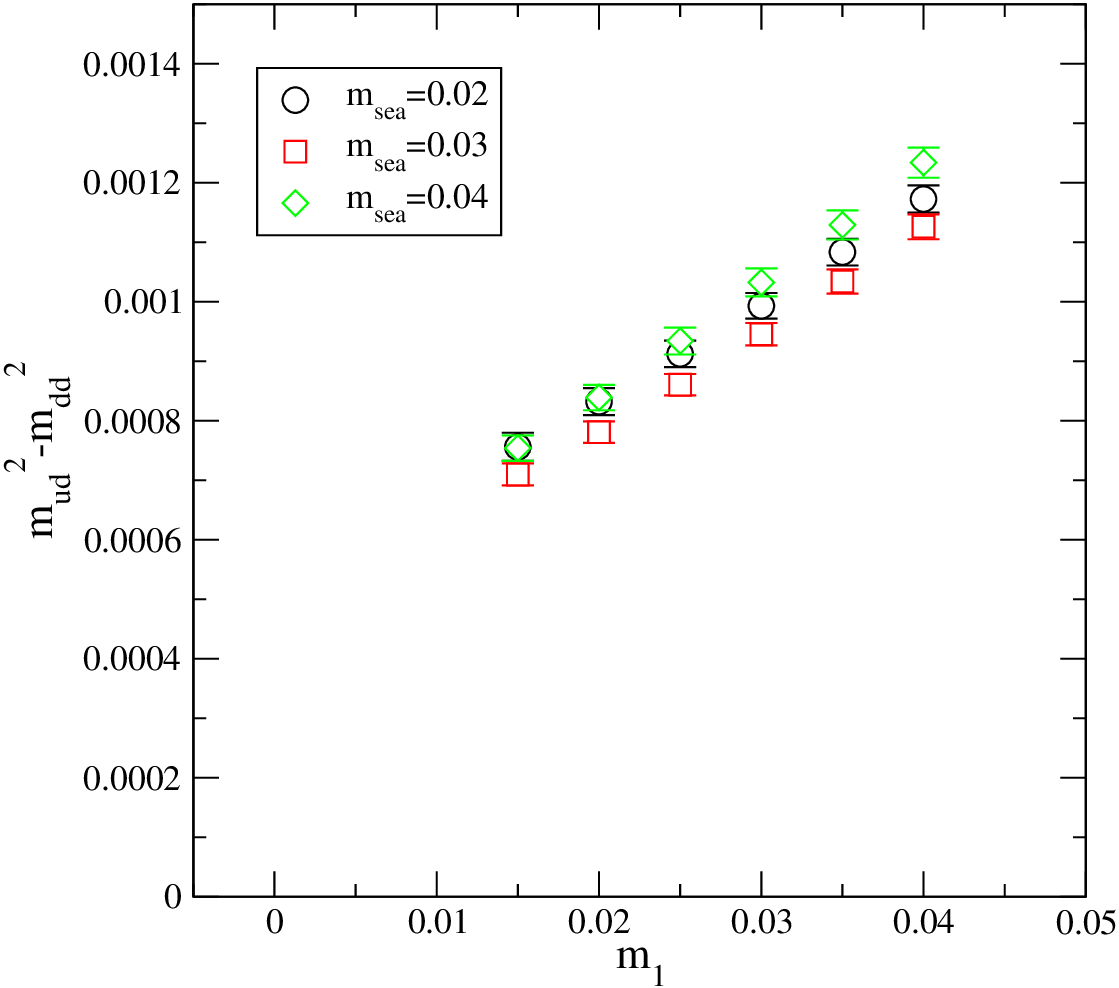}
 \caption{The pseudoscalar mass-squared splitting dependence on the sea quark mass for a representative case. Only splittings of degenerate mesons are shown.}
 \label{fig:sea quark dep}
\end{figure}
\end{document}